\documentclass[aip, chaos, reprint]{revtex4-2}
\usepackage{amsfonts,amssymb,amsmath,times}
\usepackage{graphicx}
\usepackage{bm}
\usepackage{enumerate}
\usepackage{color}
\usepackage{todonotes,hyperref,url}
\usepackage{gensymb}
\usepackage{booktabs} 

\begin{document}

\title{Hierarchical Synchronization and Distortion Scaling in Social Media Networks: A Fractal-Like Topology Theory}

\author{Kaiming Luo}
\affiliation{School of Information Science and Technology, Fudan University, Shanghai, 200438, China}
\affiliation{School of Physics and Electronic Science, East China Normal University, Shanghai, 200241, China}

\date{Received: date  \today}

\begin{abstract}
The rapid proliferation of social media as a dominant channel for information dissemination has intensified concerns over systemic information distortion—whereby content is progressively altered through successive layers of transmission. While prior studies have explored such distortion qualitatively, the quantitative interplay between propagation topology and stochastic cognitive perturbations remains insufficiently understood. In this work, we propose a novel fractal-inspired directed hierarchical network (DHN) model to capture the structural patterns of propagation, and introduce a Noise-Frustrated Hegselmann–Krause (NFHK) framework to model opinion dynamics under noise. Analytical results, supported by group and graph theory, reveal that noise accumulation leads to increasing opinion distortion and the emergence of intra-layer synchronization. Multi-agent simulations confirm these effects, showing that noise intensity shapes both convergence rates and weak intra-layer clustering. Empirical validation using a representative retweet cascade demonstrates that the proposed model reproduces real-world distortion patterns and synchronization behaviors, even without direct links. This work uncovers a unified mechanism for information distortion in digital platforms and offers topology-aware insights for public opinion governance and platform regulation.
\end{abstract}

\maketitle

\section{Introduction}

The architecture of contemporary information ecosystems has been fundamentally reshaped by social media platforms such as X, Facebook, and Weibo. While fulfilling a critical societal function of rapid information exchange, these platforms also introduce systemic vulnerabilities to nonlinear distortion, where message fidelity deteriorates progressively across transmission layers \cite{pariser2011filter, vosoughi2018spread, lewandowsky2017misinformation, friggeri2014role}. This process gives rise to two notable paradoxes: (i) even under bounded confidence thresholds in opinion dynamics, significant misalignment emerges between sources and downstream recipients \cite{hegselmann2002opinion}; and (ii) geographically dispersed user groups often exhibit spontaneous opinion synchronization without direct interaction \cite{nicosiaPRL2013}.

\begin{figure*}
    \centering
    \includegraphics[width=0.8\linewidth]{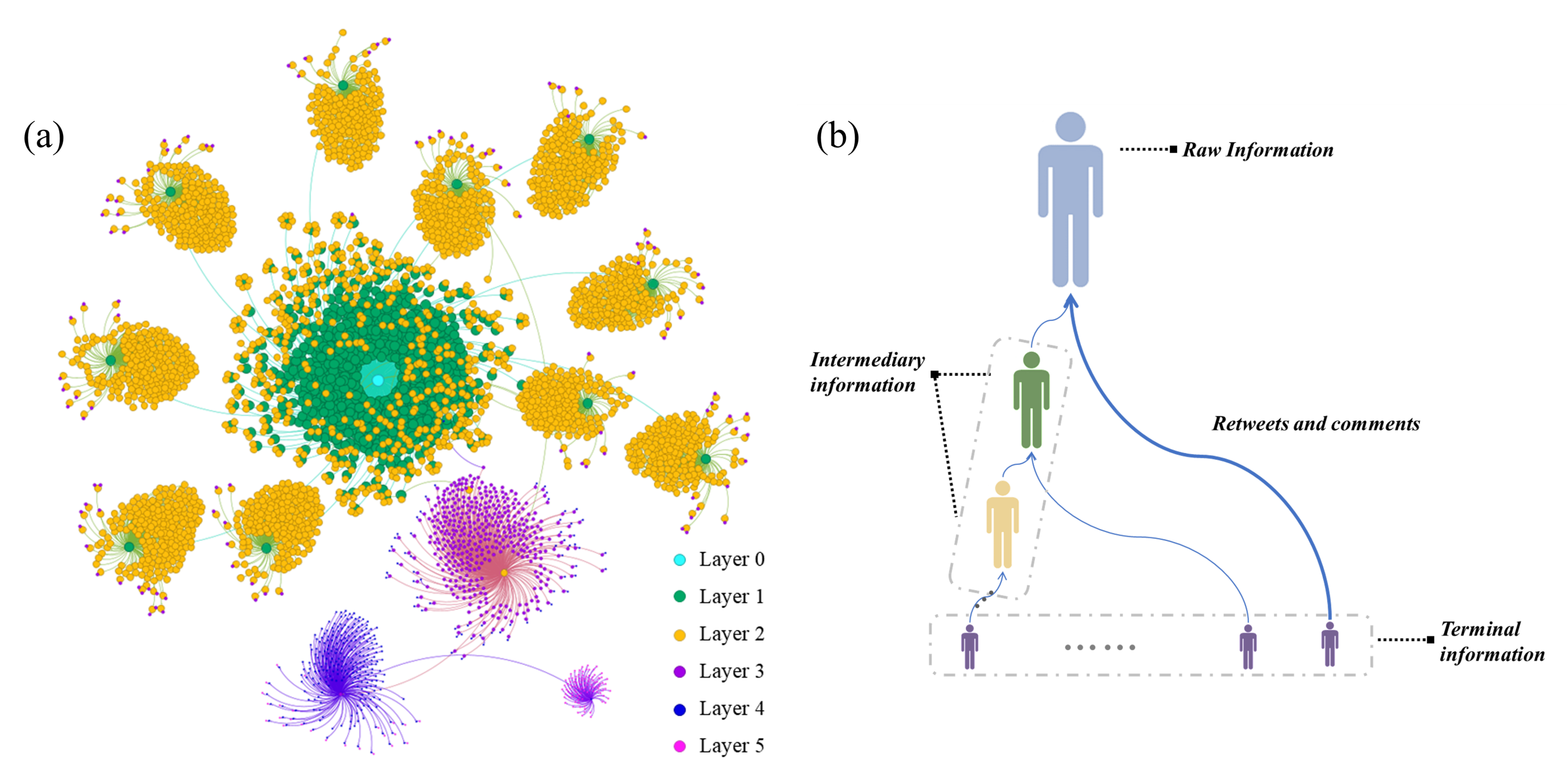}
    \caption{Schematic overview of hierarchical information propagation on social media platforms. (\textbf{a}) \emph{Network topology of retweets and comments}: Each node represents a user account, with directed edges indicating retweets or comments. The central \textit{red} node denotes the original information source. The first layer (\textit{orange}) comprises users who directly retweet from the source; the second layer (\textit{yellow}) includes users influenced by the orange-layer nodes. Peripheral \textit{pink} nodes represent end-users who receive but do not further propagate the information. A continuous color gradient encodes propagation depth.(\textbf{b}) \emph{Multi-stage content transformation}: The dashed arrow emanating from the top \textit{red} figure signifies the generation of ``Raw Information.'' Intermediate figures (\textit{orange} and \textit{light green}) generate ``Intermediary Information'' through added commentary or framing. These modified messages reach final recipients (\textit{purple} figures) as "Terminal Information". Labels denote key transitions between information states.}
    \label{fig:mechanism}
\end{figure*}

Studies of complex networks further reveal heterogeneous and structured information pathways that shape information propagation \cite{Ghavasieh2024}. As illustrated in Fig.~\ref{fig:mechanism}(b), information typically originates from authoritative sources (e.g., institutions or influencers), passes through intermediary amplifiers such as journalists and engaged citizens, and finally reaches a broad audience. At each stage, intermediaries not only relay but also reinterpret content, introducing commentary that nonlinearly amplifies distortion—a process modulated by both network structure and human cognitive biases \cite{centola2007complex, tversky1974judgment, pennycook2021psychology}. Moreover, algorithmic recommendation systems interact with social-cognitive processes to facilitate complex contagion across geographic and ideological boundaries \cite{bakshy2015exposure, nguyen2014exploring, centola2010spread, romero2011differences}.

Quantifying these distortion mechanisms is essential to model how minor content perturbations cascade through social networks, leading to belief polarization and misinformation spread \cite{lazer2018science, pennycook2021accuracy}. Systematic measurement can also inform platform-level interventions—such as algorithmic adjustments and adaptive feedback controls—to mitigate echo chambers and limit misinformation virality. Addressing these challenges thus requires an integrative framework that bridges network-scale analysis of information cascades with the nonlinear dynamics of individual belief formation.

A critical manifestation of these systemic vulnerabilities is the significant distortion inherent in real-world information dissemination. Even with limited confidence thresholds, the original viewpoint of a news item can be progressively altered as it propagates across layers, potentially leading to completely contradictory opinions \cite{bikhchandani1992information}. Social media platforms notably amplify this distortion effect \cite{vosoughi2018spread, allcott2017social}. Social-media users can aggregate into homophilic and polarized communities, forming echo chambers \cite{cinelli2021echo, delvicario2016echo}. In the hierarchical setting considered here, we investigate a distinct dynamical mechanism in which nodes at the same propagation depth can converge despite the absence of direct intra-layer links; we refer to this model behavior as hierarchical synchronization. This phenomenon resembles synchronization phenomena observed in physical systems \cite{bergnerPRE2012, Luo2024APS, Luo2024CSF,HASHEMI2025117205,ANSARINASAB2025112144}.

However, the dynamical mechanisms driving such synchronization constitute a critical gap in current opinion dynamics theory. Specifically, prevailing models face three key limitations: (1) Conventional opinion-dynamics models \cite{hegselmann2002opinion, DeGroot1974, zhang2013PhysicaA} fail to capture the scale-invariant connectivity of real social networks; (2) Information distortion is often treated as simple additive noise, neglecting its emergent nature from network-cognition interactions; and (3) The phenomenon of hierarchical synchronization among topologically distant users defies explanation by existing social influence frameworks \cite{bergnerPRE2012}. Our work addresses these gaps by introducing a dynamics-aware framework that explicitly links network hierarchy to emergent opinion formation.

In this work, we introduce a unified modeling framework that integrates network topology with opinion dynamics. This framework incorporates two core components. First, the Directed Hierarchical Network (DHN) model, depicted in Fig.~\ref{fig:mechanism}(a), captures the characteristic, scale-invariant topology of social media cascades by organizing nodes into concentric layers with directed edges and variable in-degrees. Second, the Noise-Frustrated Hegselmann-Krause (NFHK) dynamics are formulated by embedding stochastic perturbations directly into the interaction mechanism. This approach explicitly treats distortion not as mere additive noise, but as an emergent property arising from the interaction between network hierarchy and biased information flow. In concert, the DHN and NFHK establish a unified mechanism for hierarchical synchronization and distortion scaling, quantitatively linking propagation depth to the accumulation of opinion bias.

The remainder of this paper is structured as follows. Section~\ref{sec:Model_Analysis} introduces the DHN and the NFHK dynamics, forming the unified modeling framework. Section~\ref{sec:result} presents the simulation results, demonstrating hierarchical synchronization and the linear accumulation of distortion across layers. Section~\ref{sec:Group_Theory} provides a group-theoretic analysis that reveals how structural symmetry underlies intra-layer synchronization, even without direct coupling. Section~\ref{sec:Control} extends the model to control strategies, including layer-dependent noise regulation and remote synchronization management. Section~\ref{sec:case}  applies the framework to a real-world Weibo cascade, where empirical data validate the theoretical predictions and allow estimation of semantic and emotional noise intensities. Finally, Section~\ref{sec:dis} discusses the implications, assumptions, and potential extensions, and Section~\ref{sec:con} concludes by summarizing findings and outlining directions for future research

\section{Model Formulation }\label{sec:Model_Analysis}

\subsection{Directed Hierarchical network}
\label{sec:DHN}
We consider a type of directed hierarchical networks (DHNs) designed to capture the multi-layered structure of information propagation commonly observed on social media platforms. Formally, the network is represented as a directed graph \( G = (V, E) \), where nodes are partitioned into \( M+1 \) distinct layers. The top layer \( L_0 \) contains the single information source node \( v_0 \), which initiates the propagation process. Each subsequent layer \( L_i \) (\( i = 1, \dots, M \)) consists of \( N_i \) nodes. Specifically, nodes in intermediate layers (\( L_1, \dots, L_{M-1} \)) act as intermediary agents that forward, filter, or comment on information received from upper layers, whereas nodes in the bottom layer \( L_M \) serve as terminal recipients or consumers of the information. This structure allows us for modeling both the cascading transmission and the accumulation of distortion across different stages of information flow:
\[
V = \bigcup_{i=0}^{M} L_i, \quad L_0 = \{ v_0 \}, \quad L_i = \{ v_{i,1}, \dots, v_{i,N_i} \}.
\]
Edges exist only between nodes in adjacent layers, reflecting the hierarchical cascade observed in real platforms: specifically, an edge \((u,v)\) with \(u\in L_i\), \(v\in L_{i+1}\) is present with probability \(P((u,v)\in E) = p_i\), while no intra-layer edges are allowed, indicating the absence of direct interactions among same-level users during the initial propagation stage. It is important to note that, unlike a tree network where each node has exactly one parent, in DHN each node may receive inputs from multiple nodes in the previous layer, reflecting the fact that information reception in social media often aggregates multiple sources.
Specifically, the number of nodes in layer $L_i$ is determined as $N_i = \max\bigl(1, \lfloor N_{i-1} \beta_i \rfloor \bigr)$, 
where $\beta_i$ represents the scaling factor that controls the growth (or reduction) of the network width at each hierarchical level.  When $\beta_i > 1$, the network expands layer by layer, while $\beta_i < 1$ leads to contraction as depth increases. 

Notice that unlike canonical trees where each node connects to a fixed number of children and the structure forms a strict acyclic graph, social media cascades exhibit greater topological flexibility. Intermediary nodes may forward information to overlapping sets of users, and redundancy or shortcut connections often emerge, creating a more statistically self-similar or fractal-like geometry. Furthermore, the number of connections per node may vary stochastically, and the diffusion process is often shaped by heterogeneous user behaviors and attention dynamics.
\begin{figure}
    \centering
    \includegraphics[width=1\linewidth]{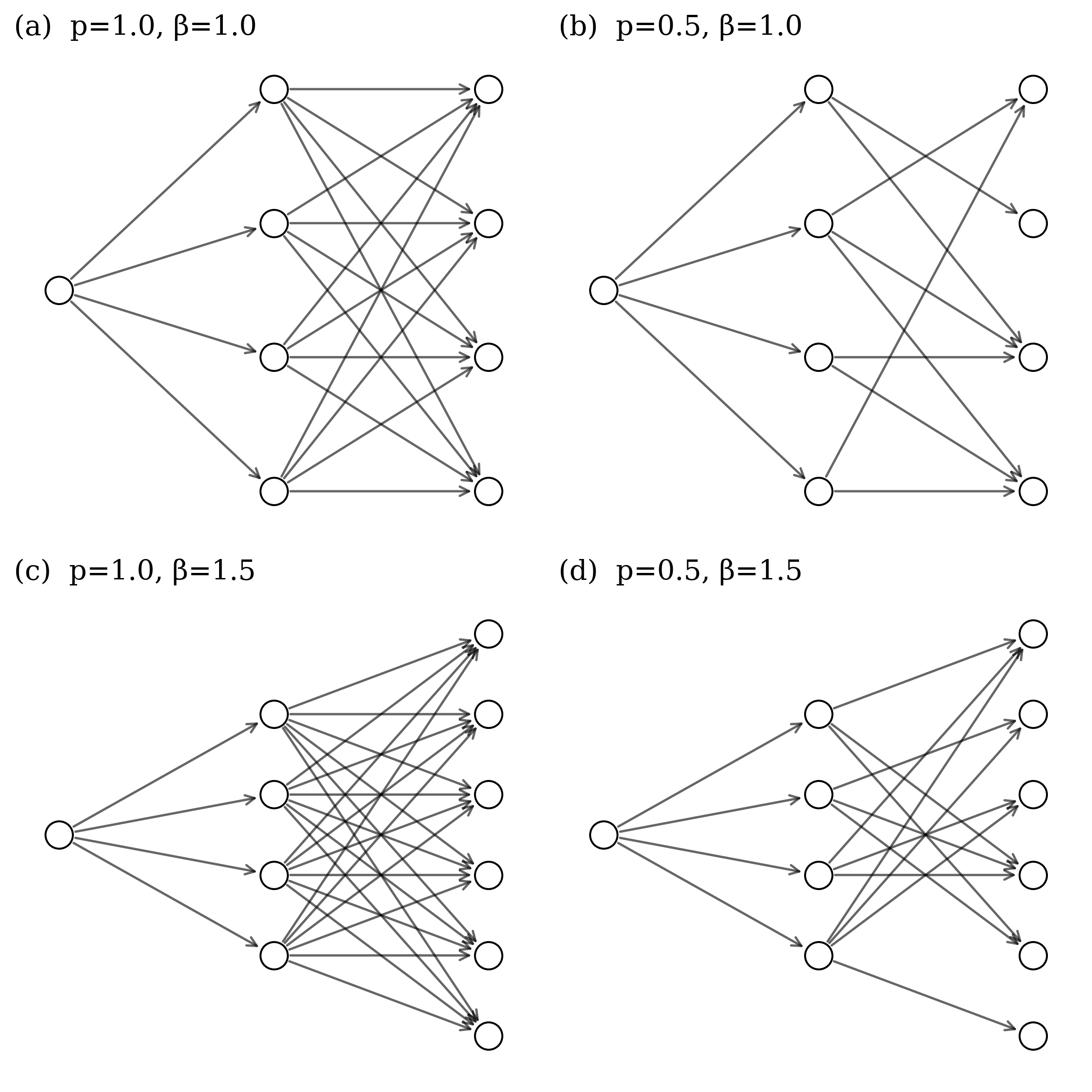}
    \caption{Example of hierarchical directed networks with varying connection probability $p$ and node scaling factor $\beta$. Each network has three layers, with one node in layer 0 and four nodes in layer 1.  The number of nodes in each subsequent layer is scaled by $\beta$ relative to the previous layer.  (a) $p = 1.0$, $\beta = 1.0$;    (b) $p = 0.5$, $\beta = 1.0$;  (c) $p = 1.0$, $\beta = 1.5$;  (d) $p = 0.5$, $\beta = 1.5$. Directed edges point from left to right, indicating the hierarchical flow of information or influence.}
    \label{fig:network_schematic}
\end{figure}

In the extreme case $p=1$, every node in layer $L_i$ connects to all $N_{i+1}$ nodes in layer $L_{i+1}$, 
yielding perfect inter-layer connectivity and strict self-similarity—in-degree distributions collapse to a delta function at $d = N_{i-1}$. 
In a more general scenario with $p=0.5$, connectivity becomes probabilistic yet still exhibits recursive branching patterns; although strict self-similarity is relaxed, the network retains the characteristic motifs of a fractal-like organization and thus maintains its influence on dynamical behavior.  The example networks shown in Fig.~\ref{fig:network_schematic} illustrate configurations where $\beta_i = \beta$ is constant across layers. 

\begin{figure}[thbp]
    \centering
    \includegraphics[width=1\linewidth]{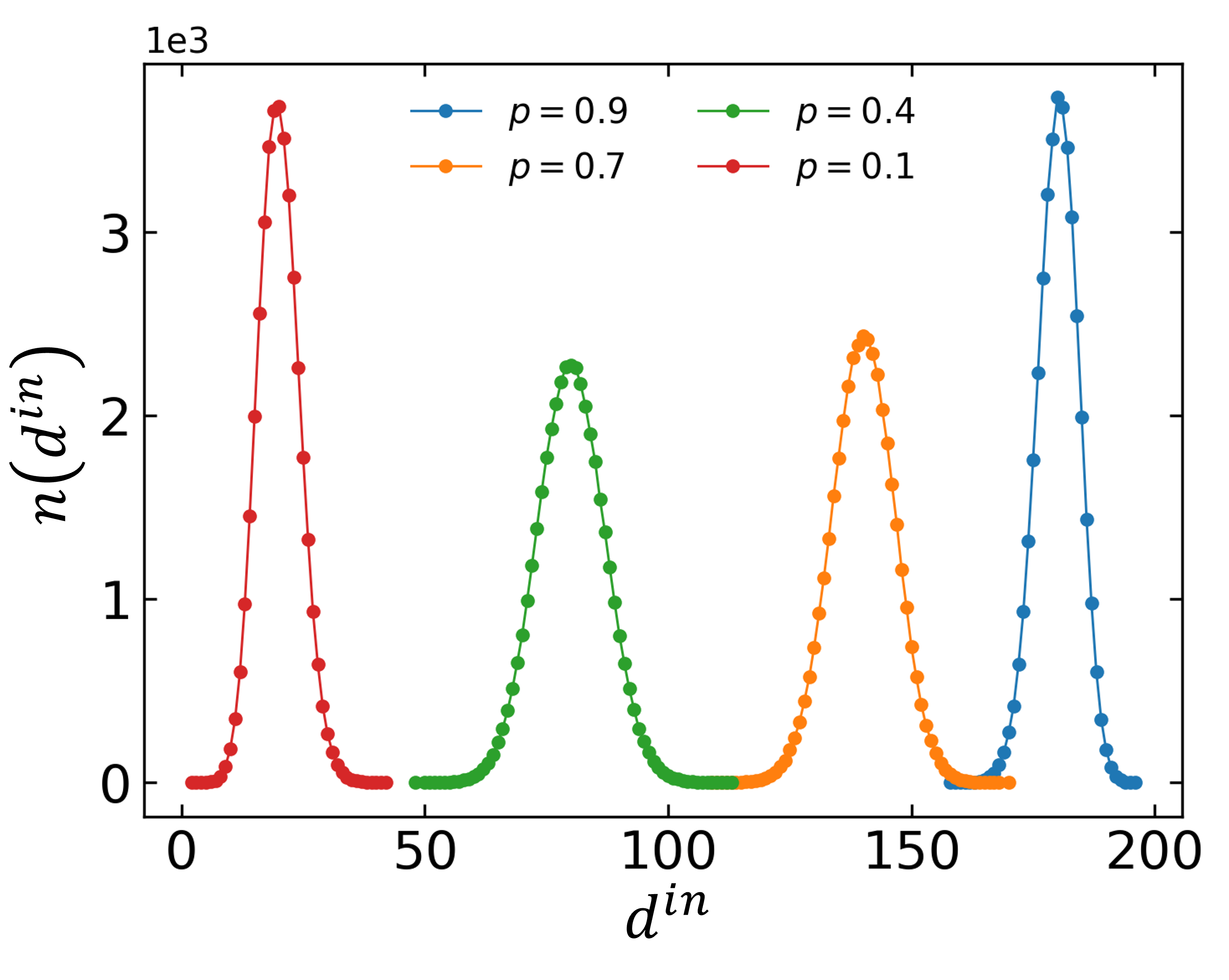}
    \caption{In-degree distributions \(n(d^{in})\) for DHN with 200 layers and 200 nodes per layer, aggregated over layers 2 to 200 to exclude trivial cases where in-degrees are fixed by construction (zero for the source node and one for the first layer). Scatter points represent numerical simulations for connection probabilities \(p=0.1, 0.3, 0.7,\) and \(0.9\), and solid lines show the corresponding theoretical binomial distributions given by Eq.~\eqref{eq:Binomial} . }
    \label{fig:degree_dist}
\end{figure}

Quantitatively, the in-degree \(d^{in}\) of a node in layer \(L_k\) follows the binomial distribution
\begin{equation}
    P(d) = \binom{N_{i-1}}{d^{in}}\,p_{i-1}^{\,d^{in}}\,(1 - p_{i-1})^{\,N_{i-1}-d^{in}},
    \label{eq:Binomial}
\end{equation}
whose most probable value in the large-\(N_{k-1}\) limit is \(d^{in}_{\mathrm{peak}} \approx \lfloor (N_{k-1} + 1)\,p_{k-1} \rfloor\). This analytical form allows direct comparison with numerical simulations. We validate the above theoretical predictions in Fig.~\ref{fig:degree_dist} by comparing numerical simulations with these analytical binomial curves: for networks of 200 layers each containing 200 nodes, statistics are aggregated over layers 2 to 200 (thus excluding the trivial in-degree of zero for the source in layer 0 and the fixed in-degree of one for layer 1). The scatter points represent simulation results for connection probabilities \(p=0.1, 0.4, 0.7,\) and \(0.9\), while the solid lines depict the theoretical binomial distributions given above (with peak at \(d^{in}_{\mathrm{peak}}\approx \lfloor (N_{k-1} + 1)\,p_{k-1}\rfloor\)). The excellent agreement across sparse to dense regimes confirms that the DHN generation process faithfully reproduces the intended scale-invariant connectivity patterns. The choice of these parameters ensures a general validation of Eq.~\eqref{eq:Binomial}, while the selected values of \(p\) cover a wide range of connection densities commonly encountered in real-world networks. In the limiting case \(p=1\), the distribution collapses to a single peak at \(d^{in}=N_{k-1}\), as expected. This validation underpins the subsequent analysis of hierarchical synchronization and distortion accumulation.

Therefore, when we consider the interaction of information in DHNs, each node at a certain layer is only influenced by the nodes at the upper layer. Taking the dynamics of the 2 layer nodes as an example, the  dynamics of the first-layer nodes are only influenced by nodes in the same layer and the information source node.

In our subsequent calculations, we assume the case of an unweighted network. However, it should be noted that all our conclusions are applicable to weighted networks as well, because obviously we can re-normalize the weights by adding appropriate nodes and edges, thereby achieving the same topology as the original network.

\subsection{Noise-Frustrated Hegselmann-Krause Dynamics}
In online social media environments, opinion dissemination often involves hierarchical forwarding combined with random disturbances. The classic Hegselmann--Krause model successfully describes bounded-confidence opinion aggregation in many settings. However, it exhibits two significant limitations when applied to social platforms. First, it neglects environmental noise such as algorithmic recommendation biases, cross-topic interference, or user attention fluctuations, even though empirical studies demonstrate that random perturbations can produce substantial opinion deviations \cite{pennycook2021psychology}. Second, its fixed confidence threshold $\varepsilon$ does not align with the broad information exposure experienced by users: on many platforms, users receive content from followers or algorithmic recommendations regardless of opinion distance, rendering a small or fixed $\varepsilon$ unrealistic and undermining the explanatory power of bounded-confidence in this context. 

To address these issues, we introduce the Noise-Frustrated Hegselmann--Krause (NFHK) model by removing the bounded-confidence constraint and embedding noise directly into the coupling term, yielding a framework better suited for opinion dynamics under social-media-like random perturbations.

We begin from a generic HK-like dynamical system in which noise is embedded in the coupling difference. The general form can be written as
\begin{equation}
  \dot{x}_j \;=\;  F_j(x_j) + \sum_{k=1}^N \kappa\,A_{jk}\,\phi(x_j, x_k) \bigl( x_k - x_j + \mu_j\,\xi_j(t) \bigr),
  \label{eq:NFHK_full}
\end{equation}
where $x_j$ denotes the opinion state (or attitude) of node $j$. The term $F_j(x_j)$ represents any intrinsic drift or external forcing acting on node $j$. The coupling sum describes how node $j$ updates its opinion by interacting with its neighbors: $\kappa$ is the coupling strength; $A_{jk}=1$ if node $k$ influences node $j$, otherwise $A_{jk}=0$; and $\phi(x_j, x_k)$ is the confidence function, which in the classic HK model equals $1$ when $\lvert x_k - x_j \rvert \le \varepsilon$ and $0$ otherwise. In the unbounded-confidence case we set $\phi(x_j, x_k) \equiv 1$. The parameter $\mu_j$ represents the noise sensitivity of node $j$, and $\xi_j(t)$ is a Gaussian noise term with mean $\xi_0 = 0.5$ and correlation $\langle \xi_j(t) \xi_j(s) \rangle = \delta(t - s)$. In the context of social media, $\xi_j(t)$ models external perturbations such as algorithmic recommendation bias, insertion of unrelated content, or fluctuations in user attention~\cite{bakshy2015exposure, lazer2018science}. The coefficient $\mu_j$ quantifies how strongly node $j$ responds to these disturbances.

Specifically, a zero-mean Gaussian noise would correspond to unbiased fluctuations arising from cognitive uncertainty or environmental randomness. However, such noise lacks the ability to induce sustained directional shifts in collective opinion and therefore cannot account for the systematic distortions observed in real-world hierarchical communication systems, where opinions often drift persistently due to external influences. By contrast, the Gaussian noise considered here, with mean $\xi_0 = 0.5$, captures persistent external biases that systematically drive opinions in specific directions, as observed when filtering algorithms, recommendation systems, or media narratives introduce consistent distortions in information exposure~\cite{DelVicario2016, vosoughi2018spread, cinelli2021echo}. The cumulative effect of this biased noise across hierarchical layers leads to systematic opinion shifts and amplifies distortions during information propagation. Importantly, the specific value of $\xi_0$ does not affect the fundamental mathematical structure of the system, as the bias can always be rescaled through an effective noise intensity according to $\mu' \xi_0' = \mu \xi_0$, where $\mu'$ and $\xi_0'$ denote the rescaled parameters.

Embedding noise via the term $\mu_j \xi_j(t)$ implies that random Gaussian fluctuations directly affect the information received by node $j$ from its neighbors. The received signal thus consists of the neighbor's opinion combined with Gaussian noise interference, reflecting how algorithmic biases, irrelevant content, or cognitive noise can distort information before integration. This formulation has two key consequences. First, moderate noise can, counterintuitively, promote or accelerate alignment of opinions by effectively enhancing coupling strength or helping the system overcome small barriers, reminiscent of noise-induced synchronization phenomena in nonlinear dynamical systems~\cite{TeramaeTanaka2004, PikovskyRosenblumKurths2001}. Second, the cumulative effect of these disturbances during repeated interactions causes the final opinion states to systematically deviate from those of a purely deterministic system, with the magnitude of deviation governed by the noise intensity $\mu$, coupling strength $\kappa$, and interaction frequency.

\begin{figure}[htbp]
    \centering
    \includegraphics[width=1\linewidth]{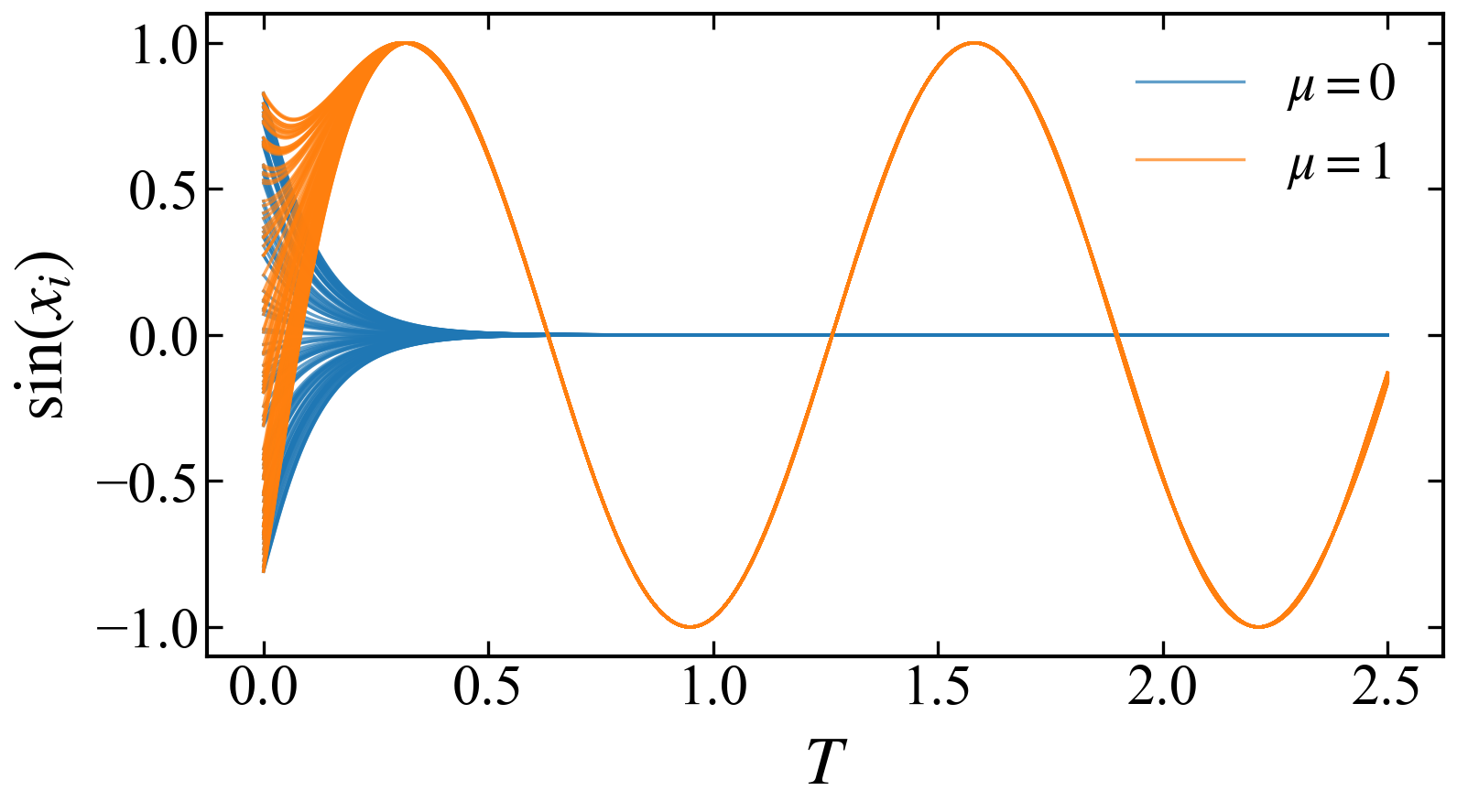}
    \caption{Time series of $\sin(x_i)$ for $N=100$ oscillators in the Noise-Frustrated Hegselmann-Krause (NFHK) model, expressed by Eq.~\eqref{eq:NFHK_simplified}. The initial phases are uniformly distributed in $[-1, 1]$ with zero mean. Blue lines correspond to the noiseless case, while orange lines represent the case with Gaussian white noise. Here, we set $\mu_1=\mu_2=\mu$.}
    \label{fig:oscillator_comparison}
\end{figure}

Based on the characteristics of online social platforms, we simplify this model. Retweeting behavior depends on heterogeneous source, message, and user-level factors \cite{ParkKaye2019, BoehmerTandoc2015, metaxas2015retweets}. To isolate the effects of topology and stochastic forcing rather than explicitly modeling these factors, we adopt the unbounded-confidence limit $\phi(x_j, x_k)\equiv 1$ as a simplifying assumption. Second, we neglect intrinsic drift $F_j(x_j)$ at this stage (setting $F_j=0$ or absorbing it into initial conditions) to focus on coupling and noise effects on relatively short to medium timescales. Thus, those simplifications lead to the simplified dynamics
\begin{equation}
  \dot{x}_j \;=\; \sum_{k=1}^N \kappa\,A_{jk} \bigl( x_k - x_j + \mu_j\,\xi_j(t) \bigr).
  \label{eq:NFHK_simplified}
\end{equation}
By rescaling time to normalize $\kappa$, we may, without loss of generality, set $\kappa=1$. Thus, Eq.~\eqref{eq:NFHK_simplified} reduces to
\begin{equation}
  \dot{x}_j = \mu_j d_j^{\rm in} \xi_j(t) - \sum_{k} L_{jk} x_k,
  \label{eq:simplified_NFHK_Laplacian}
\end{equation}
where $L = [L_{jk}]$ is the graph Laplacian, defined as $L_{jk} = d_j^{\rm in} \delta_{jk} - A_{jk}$, and $d_j^{\rm in}$ is the in-degree of node $j$.

We illustrate the impact of noise strength $\mu$ in Fig.~\ref{fig:oscillator_comparison}. In the absence of noise, the system converges to a well-defined steady state in which all oscillators maintain constant phase differences, reflecting a homogeneous collective behavior determined solely by deterministic coupling. The network topology, in this case, remains hidden in the final state because the coupling enforces uniformity across the system. However, when noise is introduced, the system can no longer settle into a static equilibrium; instead, the collective phase exhibits continuous temporal fluctuations, leading to a dynamic, time-dependent steady state. 

Furthermore, the presence of noise effectively dismantles the homogeneity imposed by deterministic interactions and exposes the influence of network topology on the system's macroscopic behavior. Physically, the noise acts as a persistent external perturbation that prevents the system from locking into uniform synchrony, allowing structural features of the network to manifest in the long-term dynamics. In larger DHNs, this persistent disturbance induces dynamic offsets between layers, with the magnitude and nature of these fluctuations governed jointly by the noise strength and the underlying network structure.

To quantify synchronization, we characterize the state of each node (denoted \(x_k \in \mathbb{R}\)) using complex numbers of unit magnitude, and adopt the classical order parameter \cite{kuramoto1984chemical,haken1983synergetics} defined as:
\begin{equation}
R \;=\; \frac{1}{N}\,\Bigl|\sum_{k=1}^{N} e^{\mathrm{j}x_k}\Bigr|,
\label{eq:order_parameter}
\end{equation}
where \(\mathrm{j}=\sqrt{-1}\).  Essentially, Eq.~\eqref{eq:order_parameter} computes the magnitude of the vector sum of complex numbers that represent node states, then normalizes this magnitude by the total number of nodes \(N\). By definition, \(R\) ranges between 0 and 1 (\(R \in [0,1]\)), which directly reflects the consistency of node states: when all node states are identical, the corresponding complex numbers align perfectly, yielding \(R=1\); in contrast, when node states are statistically scattered with no uniformity, the vector sum cancels out, resulting in \(R=0\).

In the dynamics of the NFHK model, we investigate how additive stochastic fluctuations modulate the coherence of the system. Let $\mu \ge 0$ represent the noise intensity in the update rule, with all other parameters fixed as specified in the model description. Empirically, an increase in $\mu$ accelerates the early-time ascent of coherence; however, it concurrently reduces the long-time coherence plateau, which is attributed to the persistent nature of stochastic fluctuations.  

\begin{figure}[h]
    \centering
    \includegraphics[width=0.75\linewidth]{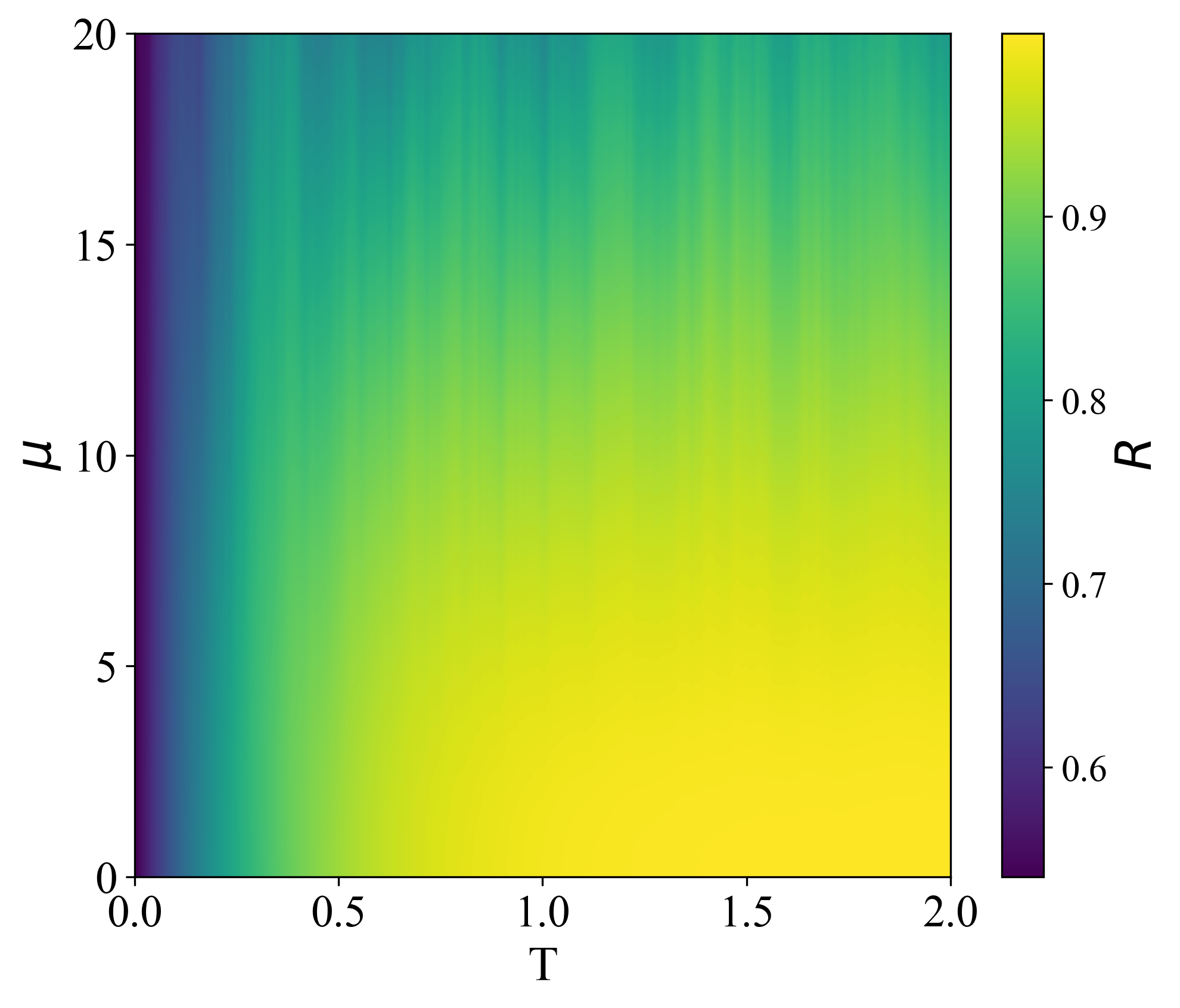}
  \caption{Noise–coherence relationship in the NFHK model, showing the global order parameter \textit{R} (Eq.~\eqref{eq:order_parameter}) as a function of time \textit{T} and noise intensity \textit{$\mu$}.}
  \label{fig:op_noise_nfhk}
\end{figure}

This characteristic behavioral pattern is quantified by the order parameter $R$, where $R$ corresponds to the expression defined in Eq.~\eqref{eq:order_parameter} when computed across all nodes in the system. As illustrated in Fig.~\ref{fig:op_noise_nfhk}, we present the temporal evolution of $R$ as a function of noise intensity $\mu$, with the results averaged over 100 independent simulation runs. Notably, smaller values of $\mu$ promote transient alignment of the system components, whereas larger $\mu$ values sustain inter-node variability and thereby diminish the asymptotic coherence level. 

For DHNs, we require order parameters that resolve coherence at different structural scales. Let layer $i$ contain $N_i$ nodes and let $S_i$ denote its index set. The layered family of order parameters is defined as
\begin{equation}
\begin{aligned}
R_i \;&=\; \frac{1}{N_i}\,\Bigl|\sum_{k\in S_i} e^{\mathrm{j}x_k}\Bigr|,\\
R_{i_1,i_2} \;&=\; \frac{1}{N_{i_1}+N_{i_2}}\,\Bigl|\sum_{k\in S_{i_1}\cup S_{i_2}} e^{\mathrm{j}x_k}\Bigr|,\\
R_G \;&=\; \frac{1}{N}\,\Bigl|\sum_{k=1}^{N} e^{\mathrm{j}x_k}\Bigr|.
\end{aligned}
\label{eq:layered_R}
\end{equation}
Here $R_i$ quantifies within-layer coherence, $R_{i_1,i_2}$ captures joint coherence across two layers, and $R_G$ measures network-level coherence. All three reduce to the same normalized resultant length on appropriate node sets, inherit the bound $[0,1]$, and increase monotonically with the concentration of $\{x_k\}$ on the unit circle. These metrics allow us to detect the typical DHN pattern: high intra-layer alignment coexisting with inter-layer offsets, where $R_i$ remains high while $R_G$ is comparatively lower.

\section{Main Result}
\label{sec:result}

\begin{figure*}[ht]
    \centering
    \includegraphics[width=1\linewidth]{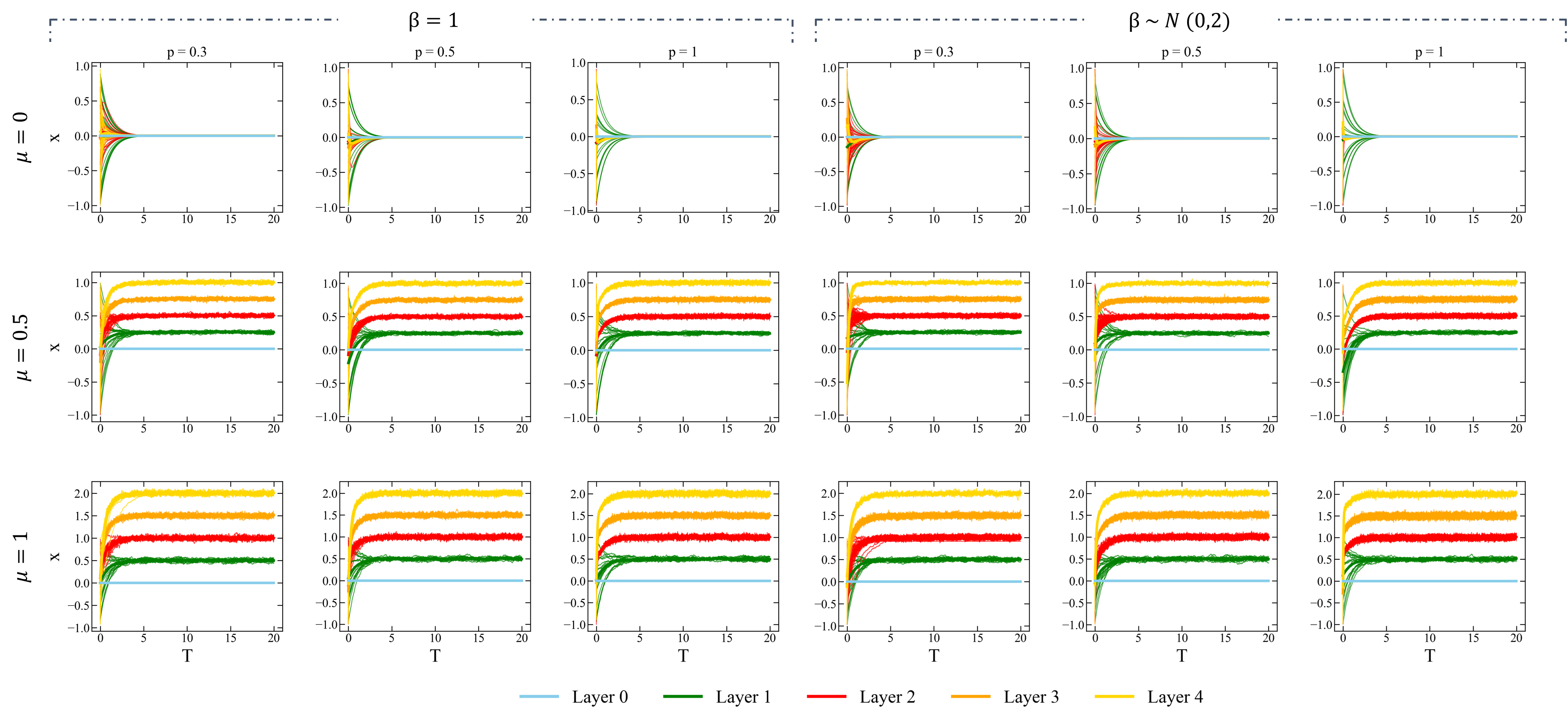}
\caption{Evolution of the $x$ over time $T$ in DHNs, governed by the NFHK in dynamics Eq.~\eqref{eq:NFHK_simplified}. The subplots denotes the system with different network parameters $(\beta,~p)$ and dynamical parameters $\mu$. The left nine subplots correspond to $\beta = 1$, and the right nine subplots correspond to $\beta \sim \mathcal{N}(0, 2)$. Each subplot shows the behavior of $x$ with lines of various colors representing nodes from different layers. }
    \label{fig:matrix}
\end{figure*}

We report the main findings from numerical simulations of the NFHK model on DHNs, as shown in Fig.~\ref{fig:matrix}. The network consists of $M=4$ layers, where layer~0 contains a single source node fixed at $x_0 = 0$, and layer~1 contains $N_1 = 20$ nodes. The mean value of the additive noise is set to $\xi_0 = 0.5$. To systematically probe the system’s dynamics, we vary the scaling factor $\beta$, inter-layer connection probability $p$, and noise strength $\mu$. Here, we first assume that $\mu_i = \mu$. Subsequently, we further discuss the more general case.

When the inter-layer connection probability is zero ($p=0$), all nodes converge to the state of the source node, $x_0 = 0$, and the system reaches a uniform steady state where $x_{i,j} = x_0$ for all nodes. In this regime, no oscillatory behavior or inter-layer differentiation is observed. The dynamics reduce to trivial convergence, illustrating that inter-layer connections are essential for generating nontrivial collective behaviors in the system.

For nonzero inter-layer connection probabilities ($p \ne 0$), a layered synchronization phenomenon emerges. Nodes within each layer evolve toward a common dynamical pattern, forming synchronized clusters characterized by small dispersion of node states around the layer mean. 

\begin{figure}
    \centering
    \includegraphics[width=0.8\linewidth]{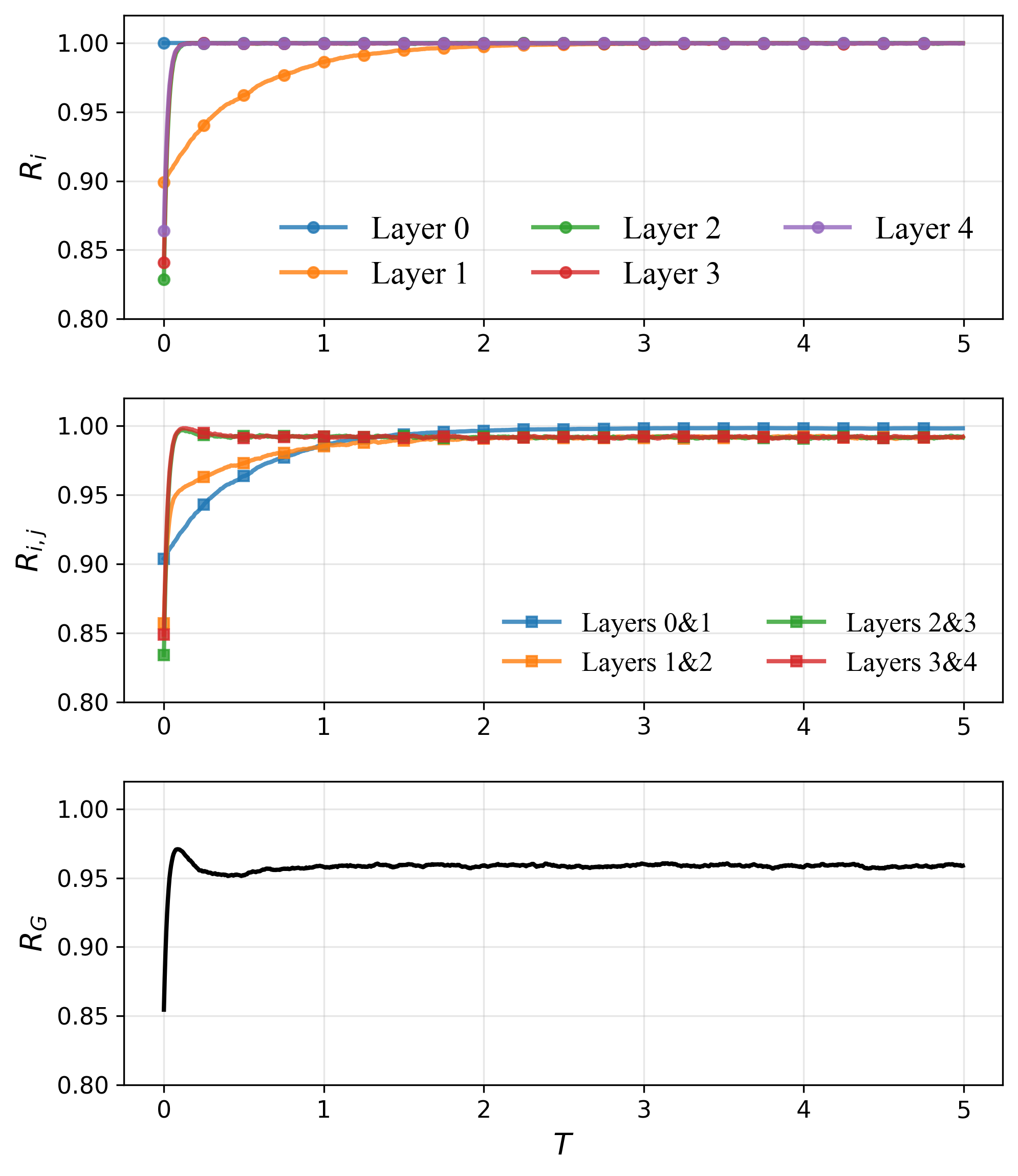}
   \caption{Time series of order parameters $R_i, ~R_{i,j},~R_G$. Parameters: $p=1$, $\beta=1$, $N_1=20$, $\mu=0.5$.}
    \label{fig:OP}
\end{figure}

We illustrate the evolution of order parameter in Fig.~\ref{fig:OP} to verify the phenomena of hierarchical synchronization, using a representative example from Fig.\ref{fig:matrix} with parameter settings $p=1$, $\beta=1$, and $\mu=0.5$. Without loss of generality, this typical parameter combination demonstrates that synchronization within each level reaches 1 in a synchronous state, while synchronization between levels remains lower than that within the levels. Ultimately, the overall synchronization degree is the weakest. This indicates that the nodes within each level achieve a synchronous state, but adjacent levels cannot reach synchronization. The system exhibits a multistable state with hierarchical dependencies, and the mismatch between the stable states leads to a decrease in system consistency. The multiple levels result in the accumulation of mismatches, which further prevents the system from achieving synchronization.

Moreover, the synchronization width within a layer can be qualitatively defined as the smallest $\sigma$ such that \(|x_{i,k} - \langle x_i \rangle| < \sigma_i\) holds for all nodes $k$ in layer $i$, where $ \langle x_i \rangle =  \frac{1}{N_i}\sum_{j=1}^Nx_{i,j} $ denotes the mean state of the layer. Simulations show that this width $\sigma$ increases with noise strength $\mu$, indicating that stronger noise leads to greater dispersion within synchronized clusters. Importantly, as $\sigma$ increases, the order parameter $R_i$ decreases accordingly. This inverse relationship reflects the fact that broader dispersion of node states reduces phase coherence, thereby lowering $R_i$. Moreover, for a fixed $\mu$, the synchronization width $\sigma$ remains nearly identical across layers, suggesting that noise exerts a uniform effect on intra-layer coherence throughout the hierarchy.

In addition to intra-layer synchronization, the system exhibits stable inter-layer state separation. The mean states of different layers display consistent offsets, forming a well-defined hierarchical pattern of synchronized clusters. The magnitude of these inter-layer differences increases systematically with noise strength $\mu$, indicating that noise not only broadens state dispersion within individual layers but also amplifies separation between successive layers. This emergence of multi-cluster synchronized states arises naturally from the network's feedforward architecture: once the system reaches a steady state, nodes in each layer act as drivers for the dynamics of the downstream layer while remaining unaffected by fluctuations originating from below. The recursive, symmetric structure ensures that each subnetwork inherits and extends the dynamic characteristics of its upstream layers, giving rise to a robust hierarchical synchronization pattern shaped by the interplay of topology, coupling, and noise.

Furthermore, the observed intra-layer synchronization constitutes a distinct form of remote synchronization within the hierarchical network. Remarkably, nodes within the same layer achieve consensus despite the absence of direct connections or mutual interactions\cite{bergnerPRE2012}. Their coordination emerges solely through the radial regulation imposed by the collective state of the preceding layer. Each lower-layer node receives input exclusively from its parent nodes in the upper layer. Although these incoming signals are statistically homogeneous and promote convergence, individual nodes typically begin with heterogeneous initial conditions and experience independent stochastic perturbations. The nontrivial nature of this phenomenon lies in the convergence of these initially diverse and noise-influenced nodes to a common synchronized state purely under shared top-down input. This emergent collective behavior highlights the striking and counterintuitive capacity of structured networks to achieve synchronization through hierarchical topology and frustrated coupling dynamics.

In the following subsections, we present a detailed quantitative analysis of these behaviors. We examine the steady-state structure of mean opinions, characterize intra-layer fluctuations, and analyze inter-layer distortion accumulation. Finally, we derive the full steady-state distribution of node opinions across layers and demonstrate that this theoretical distribution matches the numerical results with high accuracy. 

\subsection{Steady-State Opinion Structure}

Here we attempt to solve the steady-state profile.  Since all nodes in layer $i$ receive input exclusively from layer $i-1$, and the dynamics is linear, the expected opinion of any node in the same layer evolves according to the same equation. As a result, their steady-state mean opinions are equal.

The evolution of $\langle x_i \rangle$ is obtained by applying Eq.~\eqref{eq:NFHK_simplified} to the nodes in layer $i$ and averaging over time; the detailed derivation is provided in App.~\ref{sec:derivation_eq5}. This yields
\begin{equation}
\frac{d}{dt} \langle x_i \rangle = \langle d_i^{\mathrm{in}} \rangle \bigl( \langle x_{i-1} \rangle - \langle x_i \rangle \bigr) + \langle d_i^{\mathrm{in}} \rangle \mu \xi_0,
\label{eq:layer_mean_dyn_sync}
\end{equation}
which captures the interplay between alignment with upstream opinions, relaxation toward the current layer mean, and the average contribution of external noise.

In the steady state, the time derivative vanishes and Eq.~\eqref{eq:layer_mean_dyn_sync} yields
\begin{equation}
\langle x_i \rangle = \langle x_{i-1} \rangle + \mu \xi_0.
\label{eq:layer_mean_recursion_sync}
\end{equation}
Considering the recursive relation, iterating this relation starting from the source node, we obtain the steady-state mean opinion in any layer:
\begin{equation}
\langle x_i \rangle = x_0 + i \mu \xi_0.
\label{eq:layer_mean_final}
\end{equation}
Thus, the expected opinion grows linearly with layer index $i$, with slope determined by the product of noise strength $\mu$ and noise mean $\xi_0$. This result highlights the cumulative effect of noise as opinions propagate downstream in the hierarchy: each layer contributes an additive distortion independent of specific network realization  parameters $\beta$  and $p$, provided the network remains connected. 

\begin{figure}
    \centering
    \includegraphics[width=1\linewidth]{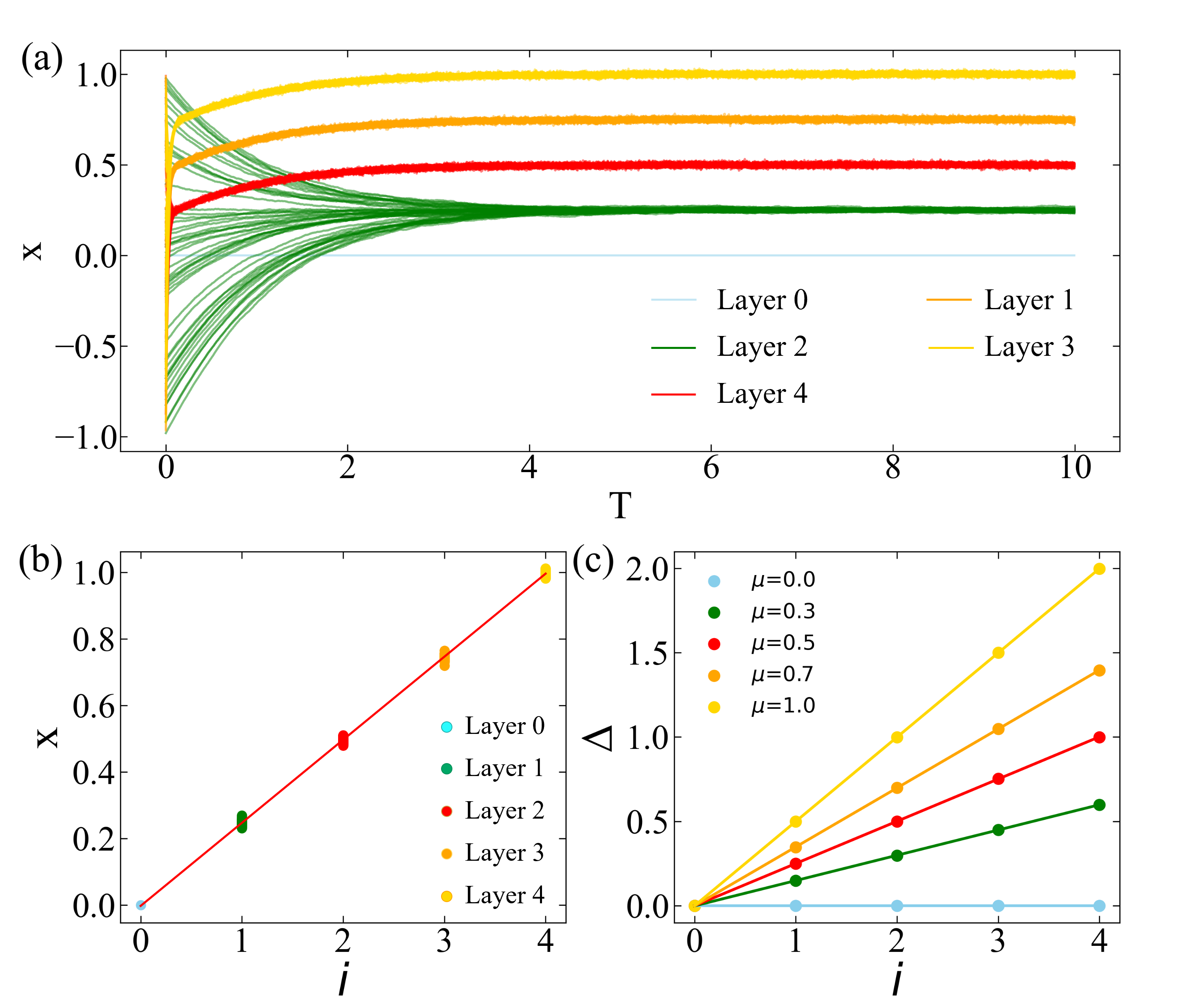}
    \caption{Results for a DHN with parameter $p = 1$,  $N = 50$. The figure illustrates the dynamic behavior and steady-state properties of nodes under these conditions. (a) Time Series of node states , where sample trajectories from each layer are depicted in different colors. ~(b) The steady-state values of all nodes as a function of their layer index, with a theoretical solution (red line) of Eq.~\eqref{eq:layer_mean_final} indicating the overall trend.~(c) The effect of varying noise strength $\mu$ on the deviation $\Delta$ between layer. The solid line represents the theoretical result in Eq.~\eqref{eq:distortion_mean}.  }
    \label{fig:steady_positions}
\end{figure}

This analytical prediction is verified in Fig.~\ref{fig:steady_positions}(c), where simulation data for various $(\mu,p)$ combinations align closely with the linear relationship of Eq.~\eqref{eq:layer_mean_final}. Even for sparse inter-layer connectivity (small $p$), the mean distortion accumulates uniformly across layers, reflecting the hierarchical structure and the additive nature of the NFHK dynamics.

In numerical simulations, the distortion $\Delta_i$ at layer $i$ is computed as
\begin{equation*}
    \Delta_i = \frac{1}{N_i}\sum_{j=1}^{N_i} \lim_{\tau \to \infty} \frac{1}{\tau - t_{s_j}} \sum_{t = t_{s_j}}^{\tau} \bigl(x_{i,j}(t) - x_0(0)\bigr),
\end{equation*}
where $t_{s_j}$ denotes the time at which node $j$ in layer $i$ reaches its synchronization threshold, and $\tau$ represents the total duration of the simulation, which is taken to be sufficiently long to approximate the long-term average. The numerical results confirm that variations in $p$ primarily affect convergence speed and fluctuation amplitude, but do not alter the linear growth of $\Delta_i$ with layer index $i$.

To make the notion of distortion explicit, we define the \emph{mean distortion} of layer $i$ relative to the source as
\begin{equation}
\Delta_i \;=\; \langle x_i \rangle - x_0 \;=\; i\,\mu\,\xi_0.
\label{eq:distortion_mean}
\end{equation}
This quantity measures the systematic bias introduced by noise after $i$ propagation steps. Eq.~\eqref{eq:distortion_mean} shows that distortion grows linearly with depth: each successive layer adds an equal expected increment $\mu\,\xi_0$.  The linear form implies that, regardless of the particular random connections (so long as the network remains connected), the average distortion is entirely determined by noise strength and layer count, not by the connection probability $p$.

A complementary perspective is to consider the inter-layer spacing between adjacent means,
\[
\Delta_i - \Delta_{i-1} = \mu\,\xi_0 \quad \text{for } i \ge 1,
\]
which reflects the uniform accumulation of noise-induced distortion at each hierarchical level. Simulation results (cf.\ Fig.~\ref{fig:steady_positions}) confirm that successive layers form equally spaced mean opinion levels, in excellent agreement with this theoretical prediction. The dependence on $\mu$ is direct: stronger noise leads to larger per-layer increments and greater overall distortion depth.

While this relation accurately captures the expected mean bias, individual realizations exhibit fluctuations around these mean positions due to finite-size effects and the randomness of connections. In subsequent sections we quantify these variances and the intra-layer spread, demonstrating that the hierarchical accumulation of distortion remains robust even in sparse networks with small $p$, where inter-layer connectivity is limited.

\subsection{Intra-Layer Synchronization and Dispersion}

In addition, we investigate the relationship between noise strength and the vertical width in a simplified discrete-time DHN model. The network consists of a single upper-layer driver node with a fixed state $x_0$ and $N$ lower-layer controlled nodes whose dynamics are governed by discrete-time linear feedback with additive noise. Specifically, the update rule for the $i$-th controlled node is given by:

\begin{equation}
    x_i(t+1) = x_i(t) + \Delta t \cdot (x_0 - x_i(t)) + \mu \cdot \xi_i(t),
\end{equation}

where $K$ is the coupling strength, $\Delta t$ is the time step size, and $\xi_i(t)$ is an independent standard Gaussian noise term at each time step, i.e., $\xi_i(t) \sim \mathcal{N}(0, 1)$. This system can be viewed as a discrete analogue of the Ornstein–Uhlenbeck process, with each controlled node subject to a stabilizing force towards the driver state and additive stochastic perturbations.

As the system evolves, the states of the controlled nodes converge in distribution to a steady state around $x_0$. Due to the linearity of the dynamics and the independence of noise, the steady-state mean of each node satisfies $\mathbb{E}[x_i] = x_0$. To characterize fluctuations within a layer, we define the node deviation $e_{i,j} = x_{i,j} - \langle x_i \rangle$. The dynamics of $e_{i,j}$ can be approximated by an Ornstein-Uhlenbeck process:
\[
\dot{e}_{i,j} = - d_j^{\mathrm{in}} e_{i,j} + \mu d_j^{\mathrm{in}} \eta_j(t),
\]
where $d_j^{\mathrm{in}}$ is the in-degree of node $j$ in layer $i$, and $\eta_j(t)$ is a zero-mean white noise process.

Solving this stochastic differential equation yields the steady-state variance:
\[
\mathrm{Var}(e_{i,j}) = \frac{\mu^2 d_j^{\mathrm{in}}}{2}.
\]
Averaging over all nodes in the layer, we obtain the intra-layer variance:
\[
\sigma_i^2 = \frac{\mu^2}{2} \langle d_i^{\mathrm{in}} \rangle,
\]
where $\langle d_i^{\mathrm{in}} \rangle$ denotes the mean in-degree in layer $i$. This result shows that fluctuations grow proportionally with both the noise strength $\mu$ and the average connectivity of the layer.

\begin{figure}
    \centering
    \includegraphics[width=0.75\linewidth]{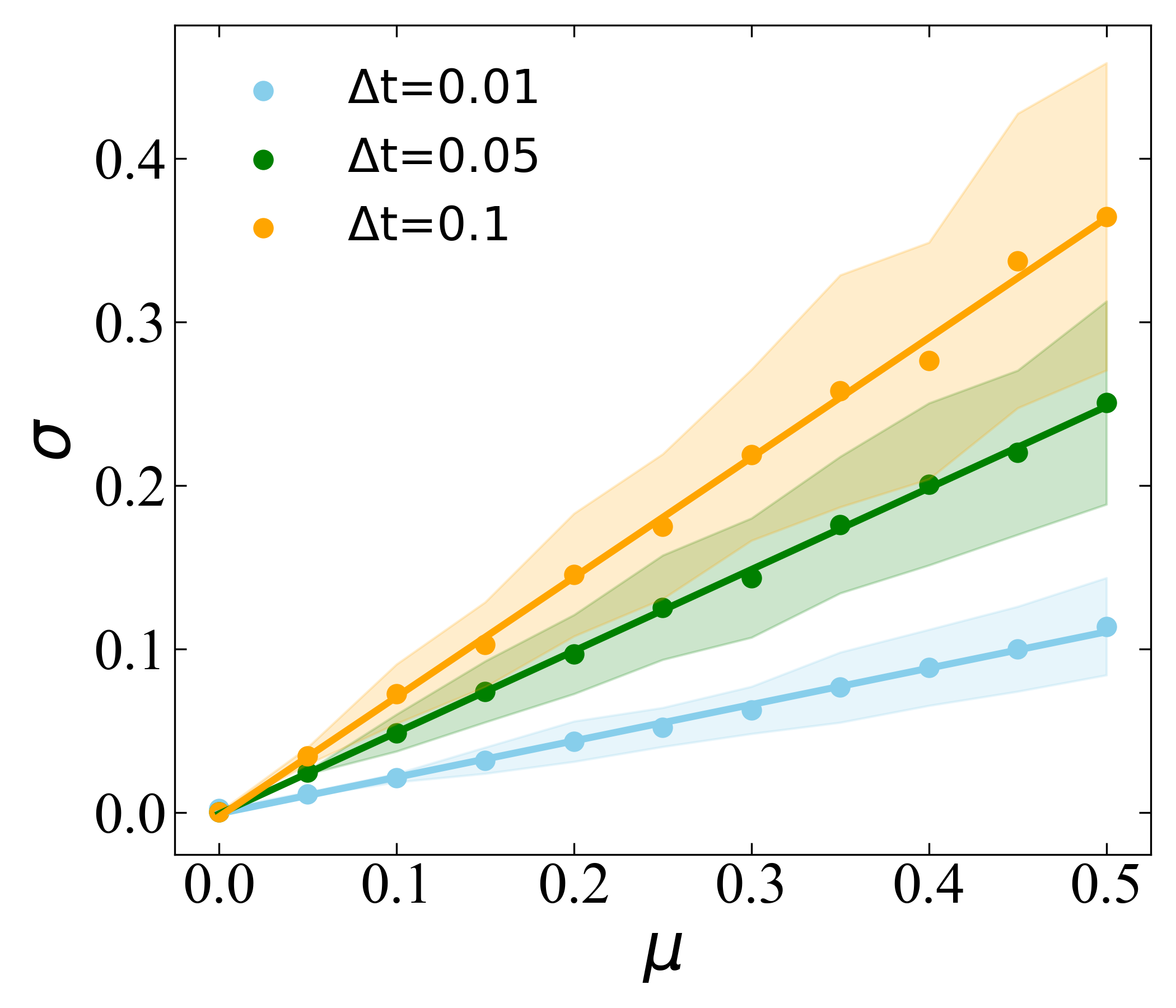}
    \caption{Relationship between noise strength $\mu$ and the vertical width $\sigma$ of phase distribution in the synchronized layer with different $\Delta t$.
    The solid line indicates a theoretical result from Eq.~\eqref{eq:width}.
    The light shaded area denotes the range of one standard deviation above and below the mean, reflecting variability across simulations.
    }
    \label{fig:intra_dispersion}
\end{figure}

To quantify the collective dispersion, we define the opinion width $\sigma$ as
\[
\sigma = \max_j x_{i,j} - \min_j x_{i,j}.
\]
By extreme value theory, its expectation scales with the standard deviation:
\begin{equation}
    \mathbb{E}[\sigma] \approx C \mu \sqrt{\langle d_i^{\mathrm{in}} \rangle},
    \label{eq:width}
\end{equation}
where $C$ is a constant depending on the number of nodes $N_i$. This theoretical prediction aligns well with numerical simulations. As shown in Fig.~\ref{fig:intra_dispersion}, when $p=1$ (i.e., fully connected inter-layer topology), the measured opinion width $\mathbb{E}[\sigma]$ exhibits a clear linear dependence on the noise strength $\mu$, in excellent agreement with the analytical result. The figure further illustrates how layer connectivity influences the magnitude of fluctuations, confirming that both noise intensity and topological density jointly determine intra-layer dispersion.

\subsection{Unified Steady-State Distribution}
\label{subsec:steady_state_distribution}

\begin{figure}[ht]
    \centering

    \includegraphics[width=0.75\linewidth]{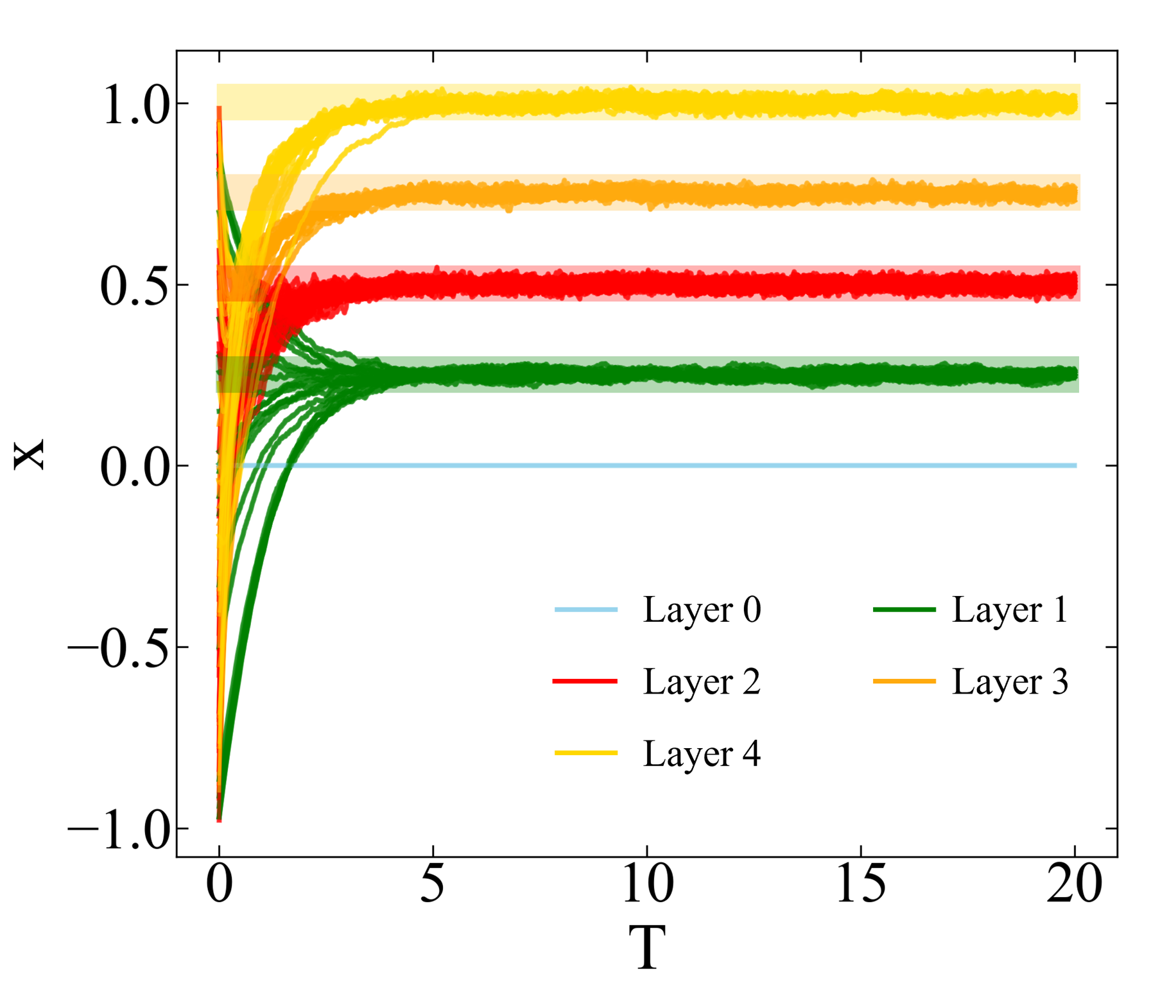}
    \caption{Time series of node states $x_{i,j}(t)$ for Layers $0–4$ in a representative NFHK simulation on a DHN. Solid lines show node trajectories: Layer 0 (light blue) remains fixed at $x_0$, while Layers $1–4$ (green, red, orange, yellow) converge toward the theoretical distribution given by Eq.~\ref{eq:Distribution_Stable} (shaded bands). Parameters:  $\Delta t = 0.01$, $N_1 = 20$, $p$ sampled uniformly from the interval $[0,1]$. Results confirm the predicted layer-wise drift and synchronization.}
    \label{fig:temporal_evolution}
\end{figure}

Combining the deterministic drift \(x_0 + i\,\mu\,\xi_0\) and the fluctuation analysis, the steady-state distribution of node opinions in layer \(i\) can be approximated by a Gaussian form:
\begin{equation}
\begin{aligned}
    &x_{i,j}(t) \sim \mathcal{N}(x_0 + i\,\mu\,\xi_0,\; \sigma^2), \quad i \ge 1,\\
    &x_{i,j}(t) = x_0, ~~~~~~~~~~~~~~~~~~~~~~~~~~~\quad i = 0,
\end{aligned}
\label{eq:Distribution_Stable}
\end{equation}
with \(\sigma^2 \approx \mu^2\langle d_i^{\mathrm{in}} \rangle/(2\Delta t)\) up to factors depending on precise noise statistics and network in-degree.Fig.~\ref{fig:temporal_evolution} compares the temporal evolution of node states obtained from numerical simulations with the theoretical predictions given by Eq.~\ref{eq:Distribution_Stable}, which shows that nodes within each layer rapidly synchronize toward a steady-state distribution centered at the predicted mean. Importantly, the simulations confirm that even in general cases where the inter-layer connection probabilities satisfy $p < 1$, our theoretical framework remains valid. The convergence dynamics and final distributions closely follow the analytical expectations, demonstrating the robustness of the derived expressions for both the layer-wise drift and intra-layer synchronization.

\section{Analytical Framework for Hierarchical Synchronization}
\label{sec:Group_Theory}

In this section we develop a theoretical understanding of hierarchical synchronization under the NFHK dynamics on DHNs. We treat two complementary regimes: first, the idealized case $p=1$, as shown in Fig.~\ref{fig:network_schematic}(a, c), where each node in layer $i$ is connected to all nodes in layer $i-1$; second, the more realistic case $p<1$ of random sparse inter-layer connections, as shown in Fig.~\ref{fig:network_schematic}(b, d).

In the case of $p=1$, the network exhibits full deterministic connectivity between successive layers, enabling exact analytical treatment through group-theoretic and automorphism-based arguments. The complete symmetry within each layer ensures that all nodes in the same layer synchronize to identical steady-state opinions, with each layer adopting a distinct value that reflects its hierarchical position and topological role.

In the case of $p<1$, the symmetry among nodes within the same layer is partially broken due to the randomness of inter-layer connections. Unlike the idealized fully connected scenario, nodes in the same layer no longer occupy identical structural roles, and their local connectivity patterns vary significantly. To overcome this challenge, we introduce a composite symmetry group that integrates both the positional information of nodes and the topology of information propagation paths. This composite group captures the residual symmetries that persist in the network despite the presence of random connections. By exploiting the invariance properties of this group, we analytically show that robust intra-layer synchronization still emerges. Specifically, the composite symmetry ensures that, although individual nodes differ locally, their dynamics are governed by statistically equivalent ensemble-averaged input patterns, which in turn generate the characteristic hierarchical synchronization observed in the system.

\subsubsection{Case for p=1}

Considering the DHN's symmetry properties enable analytical treatment of hierarchical synchronization. Consider the generalized automorphism group \( \mathrm{GAut}(G) \) of the network. For any permutation \( \pi \in \mathrm{GAut}(G) \), there exists a permutation matrix \( P_\pi \) such that:
\begin{equation}
    P_\pi L P_\pi^{-1} = L.
    \label{eq:automorphism}
\end{equation}
This symmetry implies that nodes within the same hierarchical layer \( i \) evolve identically, leading to the emergence of hierarchical synchronization.

In a connected graph, the Laplacian matrix has a zero eigenvalue, and therefore Eq.~\eqref{eq:steady_state} is singular. As a result, at each time \( t \), we can solve the system by calculating the phase difference between each node and a given reference node. Clearly, the topological features of the nodes directly determine the position of their steady-state solutions. Specifically, nodes with the same status in the topological structure converge to the same synchronized state, while nodes with different statuses converge to different steady states. Below, we provide a proof for the phenomenon of hierarchical synchronization in the context of comment/retweet networks based on group theory, specifically demonstrating that nodes within the same layer, which do not have direct interactions, synchronize their opinions.

The abstract structure of the comment/retweet network is best characterized as a fractal network, as shown in Fig.~\ref{fig:network_schematic}. Unlike traditional hierarchical models, a fractal network captures the self-replicating substructures within social media interactions, where each layer exhibits similar characteristics to the previous one. This feature allows us to apply mathematical techniques from fractal theory and group symmetry to analyze how information distortion accumulates across layers. To facilitate analysis from a topological perspective, we define this network, which can be divided into multiple sub-networks, where each sub-network exhibits similar characteristics to the original network, as a fractal network. Since fractal networks exhibit general self-similarity, the dynamic features of each sub-network are also identical. Therefore, it is sufficient to prove the dynamics of just one of the self-similar sub-networks.
\begin{align*}
    &\pi_1:\left((0,1),(1,1),(1,2)\ ,...,(1,N_{1-1}),(1,N_1)\right)\\&~~~~~~~\rightarrow\left((0,1),(1,2),(1,1)\ ,...,(1,N_{1-1}),(1,N_1)\right)\\
&\pi_2:\left((0,1),(1,1),(1,2)\ ,...,(1,N_{1-1}),(1,N_1)\right)\\&~~~~~~~\rightarrow\left((0,1),(1,3),(1,1)\ ,...,(1,N_{1-1}),(1,N_1)\right)\\
&~~~~~~~~~~~~~~~~~~~~~~~~~~~~~~~~~~~~~~~~~~~~~…\\
&\pi_{{\ N}_1!}:\left((0,1),(1,1),(1,2)\ ,...,(1,N_{1-1}),(1,N_1)\right)\\&~~~~~~~\rightarrow\left((0,1),(1,N_1),(1,N_{1-1})\ ,...(2,1),(1,1)\right).
\end{align*}

Obviously, the nodes \( (1,1), (1,2), \dots, (1,N_1) \) in the graph \( \bm{G}_{\{0,1\}} \) are symmetric, because we can relabel the nodes in \( \bm{G}^*_{\{0,1\}} \) to map the nodes in pairs, and vice versa, while the adjacency matrix of \( \bm{G}^*_{\{0,1\}} \) remains unchanged. Formally, this means there exists a permutation matrix \( P = P(\pi) \) such that \( PL_{\{0,1\}}P^{-1} = L_{\{0,1\}} \). If the permutation matrix \( P \) of the automorphism \( G \) commutes with \( L_{(j-1)~j} \), then the operation \( PL_{\{0,1\}}P^{-1} = L_{\{0,1\}} \) preserves the information in the original network adjacency matrix \( L_{(j-1)~j} \). By left-multiplying both sides of the system’s equation under synchronization conditions by the matrix \( P \), we obtain:
\begin{equation}
PL_{\{0,1\}}\bm{x}=\mu\xi P\left[\bar{k}\mathbf{1}-\bm{k}\right].
\end{equation}
Based on the commutative property of \( P \) and \( L \), we have:
\begin{equation}
L_{\{0,1\}}P\bm{x}=\mu\xi P\left[\bar{k}\mathbf{1}\ -\ \bm{k}\right].
\end{equation}
Thus, we obtain the equation for the system that exhibits exchange symmetry as:
\begin{equation}
    L_{\{0,1\}}P\bm{x}\ =L_{\{0,1\}}\bm{x},
\end{equation}
which is singular, meaning the system has only one degree of freedom. We can solve this by leaving one variable \( x \) from the \( N \) variables. Let \( \hat{x}_j = x_j - x_k \), and consider the new system:
\begin{equation}
    L_{\{0,1\}}^*P^*\hat{\bm{x}}=L_{\{0,1\}}^*x,
\end{equation}
where \( P^* \) is the matrix obtained by removing the \( k \)-th row and the \( k \)-th column from \( P \). If \( P \) does not permute the \( k \)-th node with another node, then \( P \) remains a permutation matrix. Similarly, \( L_{\{0,1\}}^* \) is the reduced Laplacian matrix, which is obtained by deleting the \( k \)-th row and the \( k \)-th column of the Laplacian matrix. By left-multiplying by \( L_{\{0,1\}}^{\ast-1} \), which is non-singular, we obtain:
\begin{equation}
    P^*\hat{\bm{x}}=\hat{\bm{x}},
    \label{con:Layer_in}
\end{equation}
Since \( P^* \hat{\bm{x}} \) represents the rearrangement of the dynamic evolution positions of the symmetric nodes, the above equation indicates that after the system reaches a steady state, the positions of the symmetric nodes converge to a single steady-state solution. Thus, the steady-state solution satisfies:
\begin{equation}
    \mathrm{diag}(x^* L) = \mu \xi \langle d^{in} \rangle I,
    \label{eq:steady_state}
\end{equation}
where \( \langle \bm{d^{in}} \rangle = \text{diag}(d^{in}_1,d^{in}_2, \dots, d^{in}_N) \) is a diagonal matrix, with each element corresponding to the degree of the respective node; and  \( x^* \) is the steady-state opinion matrix as
\begin{equation}
    x^*=
\begin{pmatrix}
\mu\xi+x_1 & x_2 & ... & x_N \\
x_1 & \mu\xi+x_2 & ... & x_N \\
\vdots & \ddots & \vdots \\
x_1 & x_2 & ... & \mu\xi+x_N
\end{pmatrix}.
\end{equation}

In summary, by leveraging the symmetry properties inherent to fractal networks and applying group-theoretic analysis, we rigorously demonstrated that hierarchical synchronization emerges naturally in DHNs when $p=1$. Nodes within the same layer, despite lacking direct interactions, converge to identical steady-state opinions due to their symmetric topological roles. Furthermore, distinct layers achieve different steady states, consistent with the structural hierarchy. These findings provide a solid theoretical foundation for understanding how topological symmetry and self-similarity drive collective dynamics in hierarchical social networks, and they offer analytical support for the numerical results presented in the main text.

\subsubsection{Case for $p<1$}
In this case, we address the synchronization properties of DHNs under the more realistic scenario where connections between layers are random rather than deterministic. Specifically, each node in layer $k$  ($k \geq 2$) receives directed edges from nodes in layer $k-1$ independently with probability $p<1$. This sparse connectivity introduces asymmetry and randomness that complicate classical symmetry-based analyses. To overcome these challenges, we introduce the concept of the \emph{intra-layer node permutation group}, which captures the statistical equivalence of nodes within each layer despite random connection patterns.

We rigorously prove that the network topology remains invariant under the action of this group, implying that the NFHK dynamics governing node opinions are equivariant with respect to intra-layer permutations. Leveraging this group equivariance and the statistical homogeneity of node inputs within layers, we derive that the system converges to a steady state exhibiting robust intra-layer synchronization. This result holds even when direct connections among nodes are sparse or missing, highlighting a fundamental synchronization mechanism driven by global layer symmetry rather than local connectivity.

To formalize the symmetry inherent in the network, we define the group
\begin{equation}
\mathcal{G} = \prod_{k=1}^{K} S_{N_k},
\end{equation}
where $S_{N_k}$ denotes the symmetric group of permutations on $N_k$ nodes in layer $k$. The group $\mathcal{G}$ acts independently within each layer, permuting node identities but not altering their layer memberships. Its action extends naturally to paths in the network: for any path $\gamma = (v_0 \to v_1 \to \cdots \to v_m)$, we define
\begin{equation}
g(\gamma) = (v_0 \to g_1(v_1) \to g_2(v_2) \to \cdots \to g_m(v_m)),
\end{equation}
where $g_k$ permutes nodes in layer $k$. 

This symmetry has a crucial topological implication. Since the connection probability $p$ and construction rules depend only on the layer indices, and not the individual node labels, any $g \in \mathcal{G}$ preserves the edge set:
\begin{equation}
(u \to v) \in \mathcal{E} \iff (g(u) \to g(v)) \in \mathcal{E}.
\label{eq:edge_symmetry_explained}
\end{equation}
This means the network topology is invariant under intra-layer permutations; in other words, two nodes in the same layer are statistically indistinguishable in terms of connection structure.

Next, consider the NFHK dynamics:
\begin{equation}
\dot{x}_i = \sum_{j \in \mathcal{N}_i^{\mathrm{in}}} \bigl( x_j - x_i + \mu \xi(t) \bigr),
\label{eq:nfhk_explicit}
\end{equation}
where $x_i$ is the opinion of node $i$, $\mu$ is the noise strength, and $\xi(t)$ is a stochastic process. The term $x_j - x_i$ represents the alignment force from neighbor $j$ to node $i$, while $\mu \xi(t)$ captures the frustration or noise in the interaction. 

We claim that the dynamics \eqref{eq:nfhk_explicit} is equivariant under $\mathcal{G}$: if $x(t)$ is a solution, so is $g x(t)$, where
\begin{equation}
(g x)_i = x_{g^{-1}(i)}.
\end{equation}
Indeed, taking the time derivative:
\begin{align}
\frac{d}{dt} (g x)_i &= \dot{x}_{g^{-1}(i)} \nonumber \\
&= \sum_{j \in \mathcal{N}^{\mathrm{in}}_{g^{-1}(i)}} \bigl( x_j - x_{g^{-1}(i)} + \mu \xi(t) \bigr).
\end{align}
By the topological invariance \eqref{eq:edge_symmetry_explained}, the in-neighbor set satisfies 
\begin{equation}
\mathcal{N}^{\mathrm{in}}_{g^{-1}(i)} = g^{-1} \mathcal{N}^{\mathrm{in}}_i.
\end{equation}
Changing summation index $j = g^{-1}(j')$ gives
\begin{align}
\frac{d}{dt} (g x)_i &= \sum_{j' \in \mathcal{N}^{\mathrm{in}}_i} \bigl( (g x)_{j'} - (g x)_i + \mu \xi(t) \bigr).
\end{align}
This shows that the transformed opinion vector $g x(t)$ satisfies the same NFHK dynamics.

We now use this symmetry to argue for layer-wise synchronization in steady state. Because of the linearity of \eqref{eq:nfhk_explicit} and the fact that each layer is statistically homogeneous due to $\mathcal{G}$, the steady-state mean opinion of any node in layer $k$ must be identical:
\begin{equation}
\langle x_a^{(k)} \rangle = \langle x_b^{(k)} \rangle, \quad \forall a,b \in V_k.
\label{eq:layer_mean_equal}
\end{equation}
In fact, the steady-state mean opinion can be written as
\begin{equation}
\langle x_{i,j} \rangle = x_0 + i \mu \xi_0,
\end{equation}
which is similar to Eq.~\ref{eq:distortion_mean}. This result follows by recursively applying the expected dynamics at each layer, since at steady state each node in layer $k$ receives statistically identical input from layer $k-1$. 

Furthermore, the uniqueness of the steady-state solution up to global shifts (under ensemble or long-time averaging) implies that any symmetry transformation $g$ leaves the solution invariant:
\begin{equation}
g x^* = x^*.
\end{equation}
This invariance forces the opinions of all nodes within the same layer to be identical:
\begin{equation}
x_a^{(k)} = x_b^{(k)}, \quad \forall a,b \in V_k,
\end{equation}
confirming robust intra-layer synchronization. Importantly, this synchronization occurs despite $p<1$, where many pairs of nodes within the same layer may not share direct connections or common neighbors. The effect arises from the global layer symmetry rather than local structural features.

To summarize, the analysis for the case $p<1$ reveals a key theoretical innovation: despite random sparse connections disrupting exact local symmetry, the concept of a composite intra-layer permutation group restores an effective global symmetry that governs the network dynamics. This symmetry ensures that the NFHK opinion dynamics remain invariant under node permutations within each layer, leading to statistically identical steady-state opinions for all nodes in the same layer. Thus, intra-layer synchronization emerges robustly from probabilistic connection structures, providing a foundational understanding of synchronization in sparse DHNs beyond fully connected idealizations.

\section{Hierarchical Synchronization Control}
\label{sec:Control}

In this section, we seek to control hierarchical synchronization by extending the original framework to a more general scenario where the noise strength varies across different layers of the DHN. Specifically, we define the noise strength as a layer-dependent list
\[
\boldsymbol{\mu} = [\mu_0, \mu_1, \mu_2, \ldots, \mu_M],
\]
where $\mu_i$ denotes the noise strength experienced by nodes in layer $i$. This setup reflects situations where different layers in the network are exposed to distinct levels of stochastic perturbations, possibly due to heterogeneous user behaviors, platform recommendation algorithms, or external environmental influences.

It is important to emphasize that this layer-dependent noise does not alter the fundamental symmetry properties of steady-state solutions of the system. As discussed in Sec.~\ref{sec:Group_Theory}, the network dynamics can still be described using the automorphism group and permutation symmetries of the DHN, ensuring that nodes within the same layer achieve statistical synchronization. However, the conclusion presented in Eq.~(16) requires generalization to incorporate the non-uniform noise strengths across layers:
\begin{equation}
\begin{aligned}
    &x_{i,j}(t) \sim \mathcal{N}\left( x_0 + \sum_{\ell=1}^i \mu_\ell \xi_0, \; \sigma_i^2 \right), \quad i \ge 1, \\
    &x_0(t) = x_0,
\end{aligned}
\label{eq:Distribution_Stable_NonUniform}
\end{equation}
where \(
\sigma_i^2 \approx \frac{\mu_i^2}{2 \Delta t}.\) This formulation captures the cumulative distortion effects introduced by successive layers, each contributing its own characteristic noise-induced shift.

\begin{figure}
    \centering
    \includegraphics[width=1\linewidth]{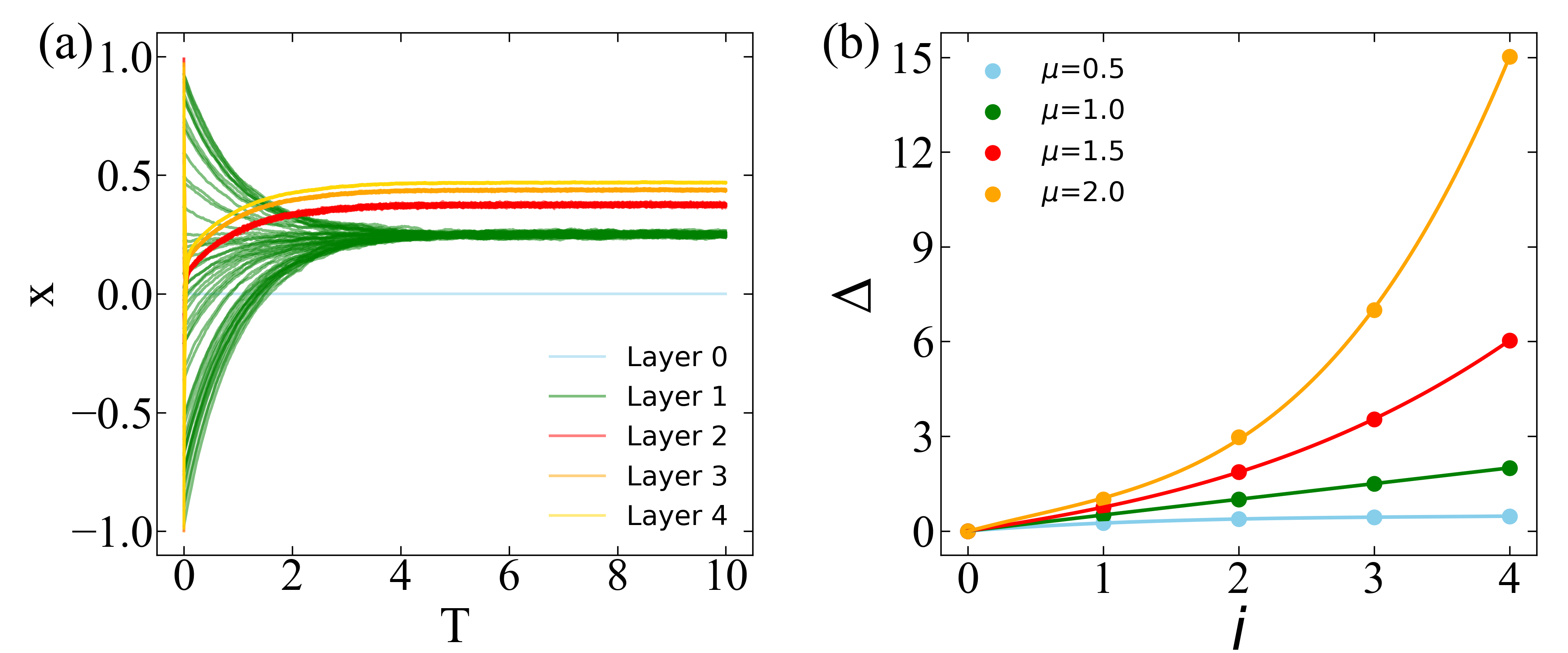}
    \caption{Steady-state opinion distribution in DHNs with non-uniform noise strengths. (a) Trajectories over time for each layer. (b) Under different $\mu$, mean deviation $\Delta$ of each layer from the source node, compared to the theoretical quadratic fit.}
    \label{fig:NonUniformNoise}
\end{figure}

We validated this theoretical prediction through numerical simulations, illustrated in Fig.~\ref{fig:NonUniformNoise}(a). The network used in these simulations contains $M=4$ layers, and the noise strengths follow the exponentially decaying pattern \(\mu_i = \alpha^{-i},\) where $\alpha=2$ is the noise scaling factor. Therefore, the specific noise strength vector is \(\boldsymbol{\mu} = [1, 0.5, 0.25, 0.125, 0.0625].\) As shown in the figure, nodes in each layer converge to distinct steady-state distributions, with hierarchical synchronization achieved within layers despite the non-uniform noise. Notably, in contrast to the uniform noise case presented in Fig.~\ref{fig:NonUniformNoise}, the offsets between adjacent layer means are no longer equal. Instead, the inter-layer differences reflect the heterogeneity in $\mu_i$.

Furthermore, we plot the steady-state mean deviation of each layer relative to the source node, demonstrating that these deviations follow a quadratic trend as predicted by the extended theoretical model. Fig.~\ref{fig:NonUniformNoise}(b) displays both the simulation results and the fitted quadratic curves, confirming the validity of our analytical framework. We also explore different values of $\alpha$ and observe consistent agreement between theory and simulation across varying noise decay rates.

This extended model provides deeper insight into how heterogeneous noise profiles shape hierarchical synchronization and distortion accumulation in social media-like networks. It highlights that controlling the noise strength at specific layers could be a viable strategy for mitigating overall information distortion in practical applications.

\subsection{Critical Conditions for Hierarchical Separation}

The emergence and regulation of hierarchical synchronization in DHNs fundamentally arise from the intricate interplay between network topology, noise characteristics, and dynamical coupling mechanisms. To achieve and maintain distinct synchronized clusters across different layers, the mean opinion gap between adjacent layers must sufficiently exceed the characteristic intra-layer fluctuations.

According to Eq.~\ref{eq:Distribution_Stable}, the difference in mean states between layers \(i\) and \(i-1\) is given by
\[
\Delta_i - \Delta_{i-1} = \mu \xi_0,
\]
where \(\mu\) denotes the noise intensity and \(\xi_0\) represents the noise bias. Meanwhile, the steady-state intra-layer standard deviation is rigorously expressed as
\[
\sigma_i = \mu \sqrt{\frac{\langle d_i^{\mathrm{in}} \rangle}{2}},
\]
with \(\langle d_i^{\mathrm{in}} \rangle\) being the average in-degree of layer \(i\), thereby encoding the influence of the network's connectivity structure on fluctuations.

For hierarchical layers to remain distinctly separated, the inter-layer mean gap must dominate over these intra-layer fluctuations. This requirement leads to the critical separation condition
\begin{equation}
|\xi_0| > \kappa \sqrt{\frac{\langle d_i^{\mathrm{in}} \rangle}{2}},
\label{eq:corrected_separation}
\end{equation}
where \(\kappa\) is a confidence parameter that quantifies the degree of statistical separation—for example, \(\kappa=2\) corresponds approximately to a 95\% confidence level assuming Gaussian noise. Importantly, this condition is dimensionally consistent and highlights that both the systematic noise bias \(\xi_0\) and the network’s average in-degree \(\langle d_i^{\mathrm{in}} \rangle\) jointly govern the feasibility of achieving hierarchical separation.

It is essential to recognize that increasing the noise intensity \(\mu\) simultaneously enlarges both the inter-layer gap and intra-layer fluctuations, making \(\mu\) a parameter with conflicting effects that cannot independently optimize both synchronization quality and layer separation. Moreover, the noise bias \(\xi_0\), reflecting systematic opinion tendencies, generally lies beyond direct manipulation by platform algorithms.

Therefore, practical and effective control strategies should focus on managing network topology and mitigating noise bias. Topology control entails reducing the average in-degree \(\langle d_i^{\mathrm{in}} \rangle = N_{i-1} p_{i-1}\) by, for instance, limiting probabilities \(p_{i-1}\). This approach suppresses intra-layer fluctuations \(\sigma_i\) and thereby facilitates clearer hierarchical separation. Noise bias correction algorithms aim to diminish \(|\xi_0|\) by filtering out systematic skewness in opinions, albeit potentially at the expense of reduced layer distinctness.

The previously held view that hierarchical separation can be straightforwardly controlled by adjusting \(\mu\) or unqualified modifications of \(\xi_0\) is therefore physically untenable and should be reconsidered. Additionally, earlier interpretations of remote hierarchical synchronization (RHS) based on alternating noise signs have been superseded by the refined RHS framework developed in Sec.~V.C, and thus are not further discussed here.

\begin{figure}
    \centering
    \includegraphics[width=1\linewidth]{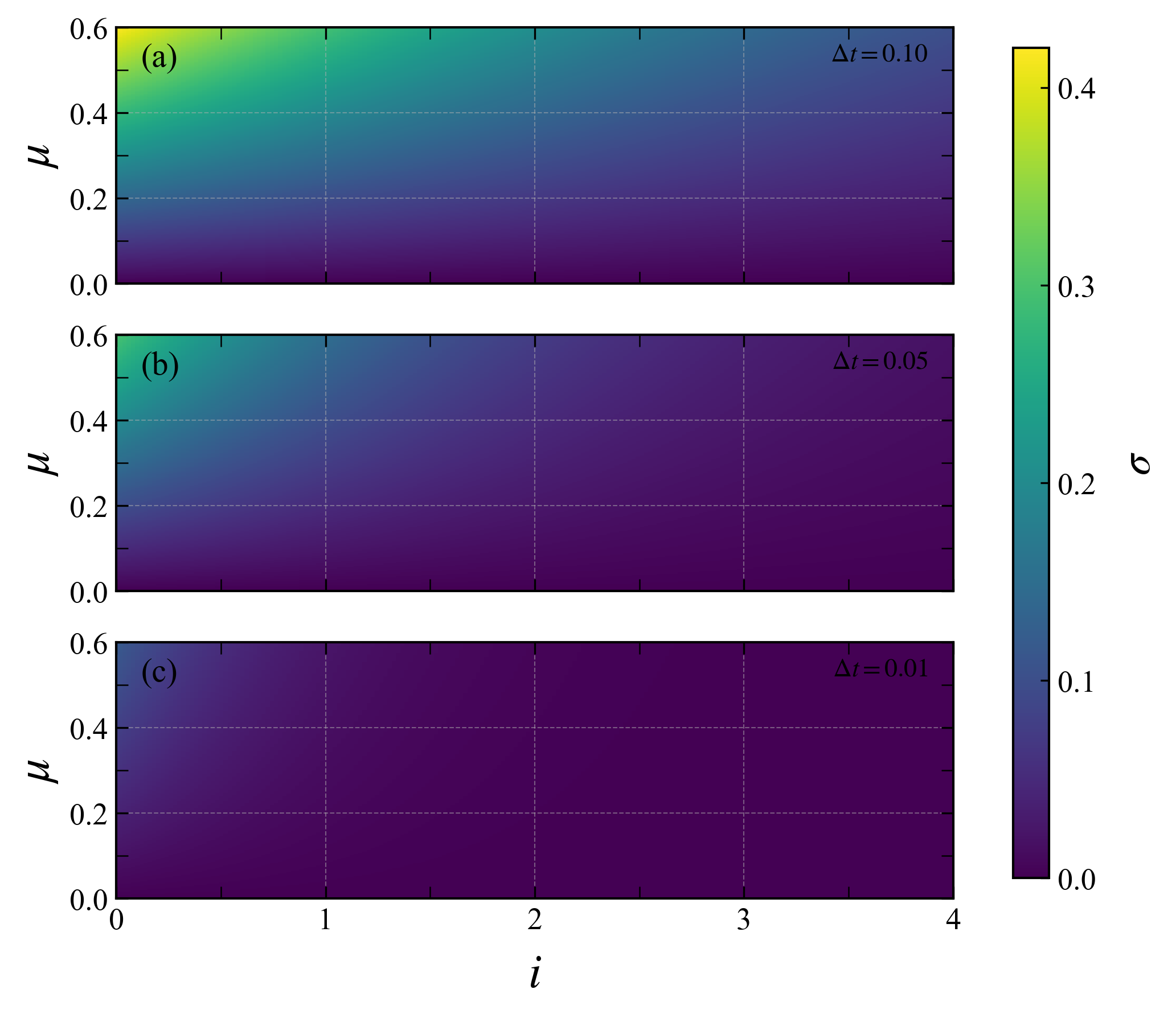}
    \caption{ Widths of each hierarchical layer for $\alpha = 0.5$. The diagrams illustrate the variation of $\sigma$ as a function of $\mu$ and $i$ under different values of $\Delta t$. The colorbar indicates the magnitude of $\sigma$.}
    \label{fig:adj_sync}
\end{figure}

Fig.~\ref{fig:adj_sync} presents the widths of node state distributions across hierarchical layers under varying conditions, providing quantitative insight into how intra-layer fluctuations depend on noise strength and dynamical parameters. Specifically, the figure displays the standard deviation $\sigma_i$ for each layer $i$ as a function of the noise intensity $\mu$ and the layer index $i$, with results shown for different values of the update rate $\Delta t$. The colorbar encodes the magnitude of $\sigma_i$, thereby enabling direct visualization of fluctuation intensities across the parameter space.

The diagrams clearly demonstrate that intra-layer fluctuations grow with increasing noise intensity $\mu$, consistent with the theoretical relation $\sigma_i = \mu \sqrt{\langle d_i^{\mathrm{in}} \rangle / 2}$. Furthermore, the effect of the update rate $\Delta t$ is evident: smaller $\Delta t$ values, corresponding to finer temporal resolution, result in reduced fluctuations, while larger $\Delta t$ amplify the dispersion. Across all cases, higher-index layers tend to exhibit larger $\sigma_i$, reflecting the accumulation of upstream noise and the amplification effect due to network depth. This systematic dependence on $\mu$, $\Delta t$, and layer index confirms the critical role of these parameters in shaping the hierarchical synchronization structure.

\subsection{Mechanism and Control of Remote Hierarchical Synchronization}

In addition to synchronization between adjacent layers, our model reveals the phenomenon of remote hierarchical synchronization (RHS), where non-adjacent layers in the DHNs exhibit identical or nearly identical mean states, while the intermediate layers remain desynchronized. This phenomenon arises due to a precise cancellation of cumulative noise effects along the path between these layers, rather than the previously assumed simple sign alternation of noise strengths.

\begin{figure}
    \centering
    \includegraphics[width=1\linewidth]{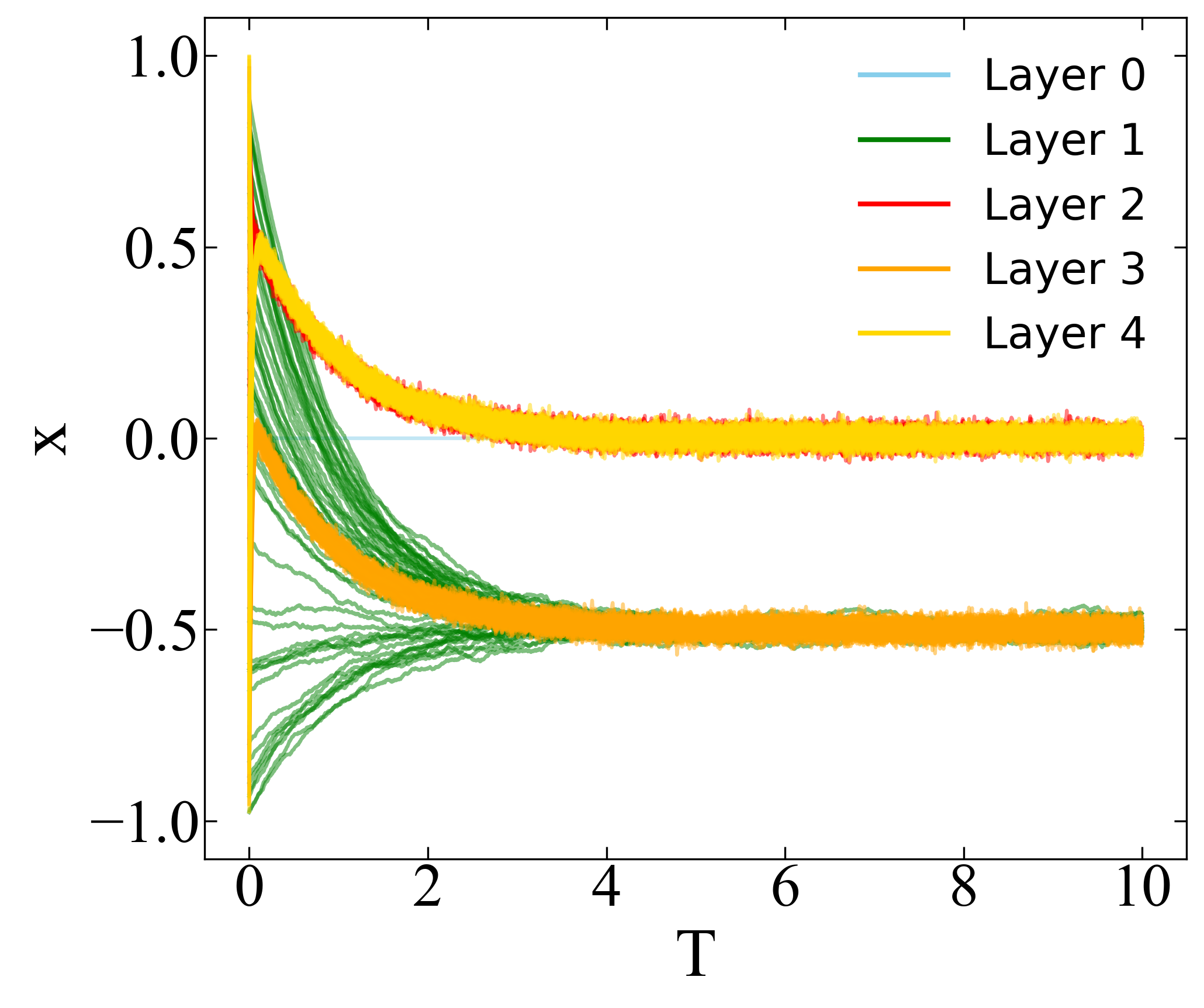}
    \caption{Demonstration of RHS in a 5‐layer cascade: despite the absence of any direct coupling between Layers 1 and 3 (Layers 0, 2 and 4), these two layers attain a common synchronized state, whereas Layer 2 (Layers 1 and 3) remains dynamically distinct due to the alternating noise intensity applied across successive propagation depths.}
    \label{fig:rhs}
\end{figure}
Consider a noise strength profile \(\boldsymbol{\mu} = [\mu_1, \mu_2, \ldots, \mu_M]^T\), with an associated noise bias \(\xi_0\). The mean state of layer \(i\) can be expressed as
\[
\langle x_i \rangle = x_0 + \xi_0 \sum_{k=1}^i \mu_k,
\]
where \(x_0\) is the fixed state of the source node (layer 0). Remote synchronization between two layers \(i\) and \(j\) (\(i < j\)) requires that their mean states coincide, which is equivalent to the \emph{cumulative noise cancellation condition}:
\[
\sum_{k=i+1}^j \mu_k = 0.
\]
This condition ensures that the noise-induced distortion accumulated between layers \(i\) and \(j\) exactly cancels out, enabling
\[
\langle x_i \rangle = \langle x_j \rangle,
\]
despite the absence of direct or adjacent coupling between them.

\begin{figure}[htbp]
    \centering
    \includegraphics[width=1\linewidth]{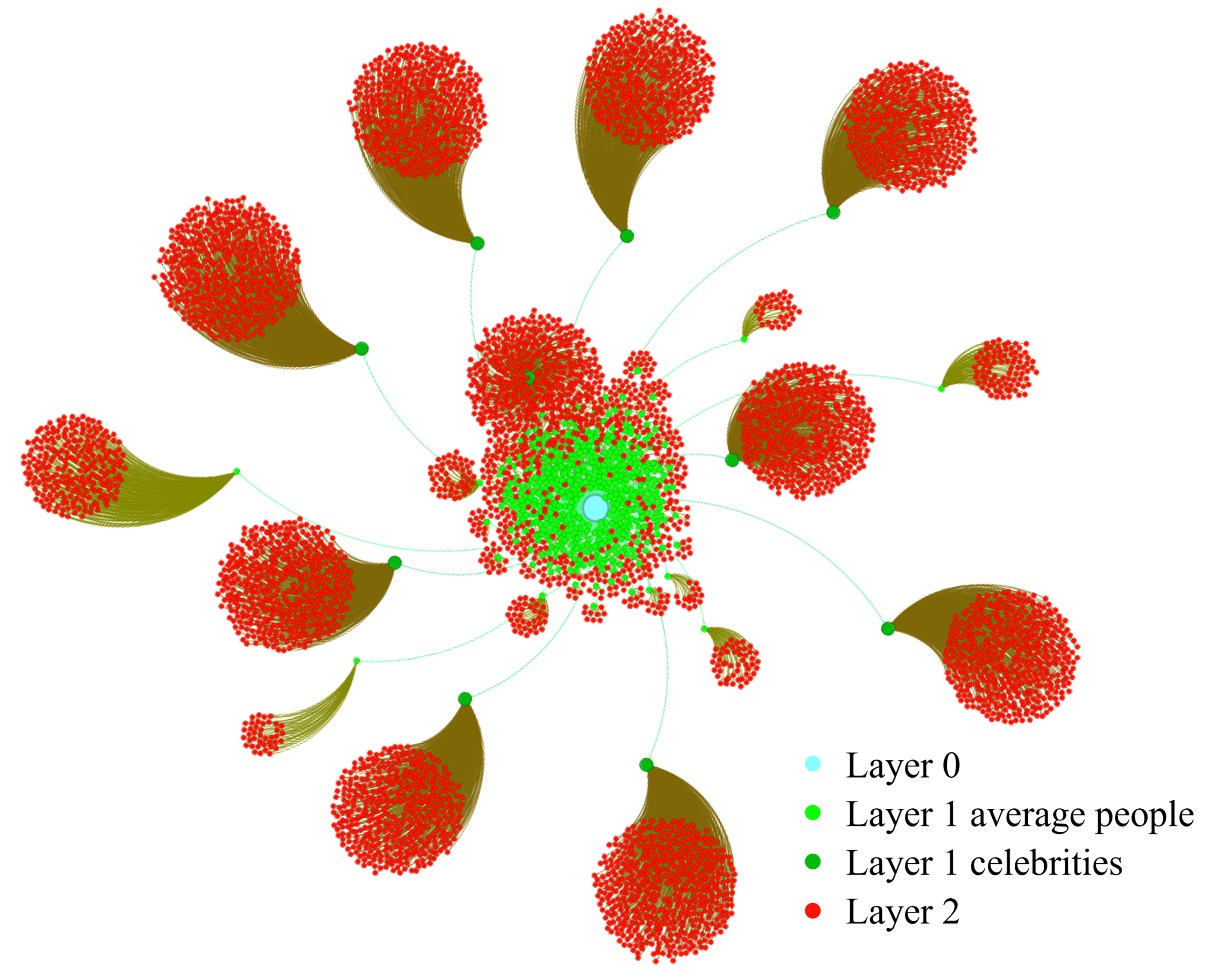}
    \caption{Visualization of a three‐layer Weibo retweet network centered on a viral CCTV News post. The original post (Layer 0, sky‐blue node) was published by the official CCTV News account and garnered over 1.31 million shares, 112 000 likes, and 6 940 comments. Layer 1 (green nodes) comprises all users who directly retweeted the source, including both high‐degree celebrity accounts (dark green, larger size) and ordinary users (light green, smaller size). Layer 2 (red nodes) represents second‐order retweeters who forwarded content from Layer 1. Node size is proportional to out‐degree (number of retweets received), and color intensity distinguishes user type and connectivity: darker hues denote higher centrality within the cascade. For visual clarity, low‐degree nodes and infrequent edges have been omitted. Detailed network parameters are provided in Tab.~\ref{tab:network_params}.}
    \label{fig:WeiboNetwork}
\end{figure}

An illustrative example of remote hierarchical synchronization is shown in Fig.~\ref{fig:rhs}, where the noise strength sequence is chosen as \(\boldsymbol{\mu} = [1, -1, 1, -1, 1]\), corresponding to noise scaling factor \(\alpha = -1\). This configuration ensures cumulative noise cancellation between certain layers, specifically \(\sum_{k=2}^3 \mu_k = 0\) and \(\sum_{k=4}^5 \mu_k = 0\), which leads to identical mean states \(\langle x_1 \rangle = \langle x_3 \rangle = \langle x_5 \rangle = x_0 + \xi_0\), while intermediate layers 2 and 4 exhibit distinct mean values.

Physically, the sign and magnitude of \(\mu_k\) encode the role of noise at each layer: positive \(\mu_k\) amplifies neighbor influence and promotes alignment, whereas negative \(\mu_k\) partially suppresses or filters upstream distortion—modeling skepticism or information attenuation—without reversing the direction of influence. Thus, negative noise strengths correspond to distortion attenuation rather than an inversion of coupling, which would require modifying the underlying dynamics. Fig.~\ref{fig:rhs} schematically illustrates this remote synchronization pattern, where non-adjacent layers share identical mean states due to the precise cancellation of cumulative noise effects. This phenomenon highlights the critical role of noise profile design in controlling hierarchical synchronization, with implications for engineering or mitigating opinion alignment across distant communities in social networks~\cite{vicario2019polarization, gleeson2016effects}.

\section{Noise Intensity Assessment}\label{sec:case}
In the social media landscape, information propagates at extraordinary speed and through intricate, multi‐tiered networks. External perturbations—or “noise”—not only degrade the fidelity of genuine content but can also exacerbate emotional polarization and collective misperception. By empirically quantifying noise intensity along these propagation chains, we can uncover how deviations accumulate across successive tiers of reception and retransmission, thereby furnishing vital data to curb the spread of misinformation, refine algorithmic recommendation systems, and develop more robust network‐synchronization strategies.

\begin{figure*}[htbp]
    \centering
    \includegraphics[width=1\linewidth]{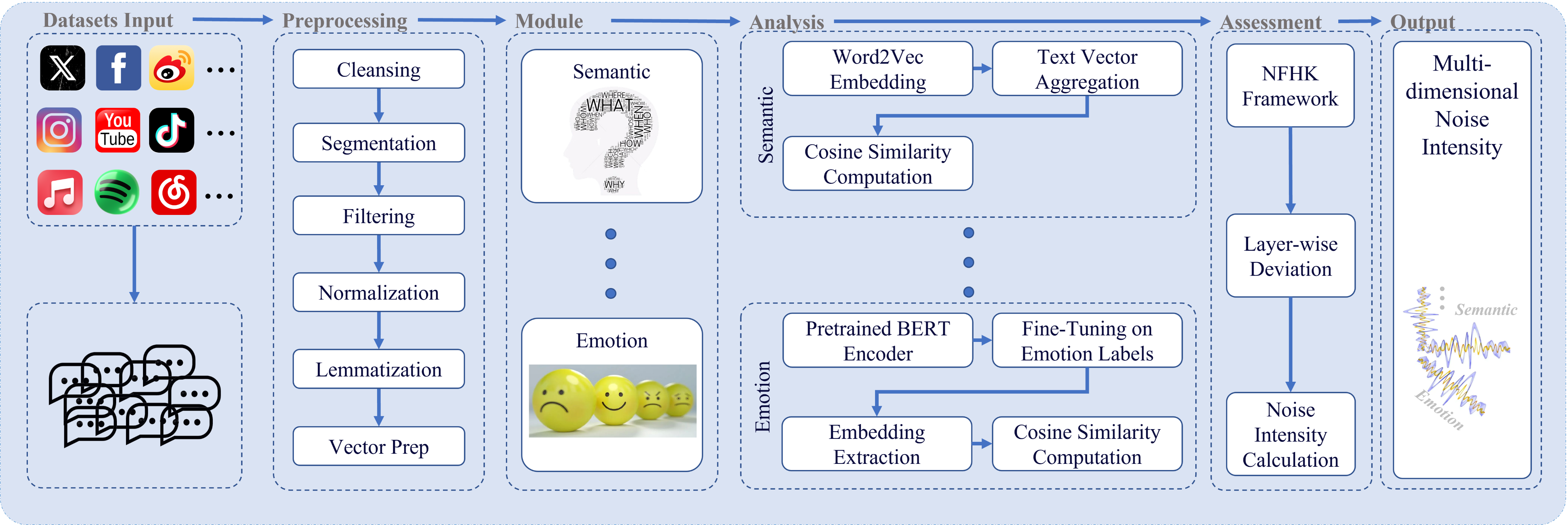}
    \caption{Overview of the multi‐channel Noise-Frustrated Hegselmann–Krause (NFHK) framework for empirical noise intensity assessment in hierarchical social media cascades. The pipeline comprises four main stages: (1) \textbf{Data acquisition and layer assignment}: raw comment or retweet logs are arranged into discrete propagation layers \(i=0,1,2,\dots\) based on their hop distance from the original source post; (2.1) \textbf{Semantic channel processing}: all text is pre‐processed to remove URLs, mentions, and non‐text tokens, then tokenized via Jieba and embedded using a Word2Vec model trained on a large‐scale Chinese corpus. Pairwise cosine similarities between layer‐wise comment embeddings yield a semantic similarity matrix \(\mathbf{C}^S\in\mathbb{R}^{L\times L}\); (2.2) \textbf{Emotional channel processing}: the same comments are fed into a fine‐tuned BERT classifier to project each text unit onto an 8‐dimensional Plutchik emotion wheel, producing emotion vectors whose pairwise cosine similarities form the emotional similarity matrix \(\mathbf{C}^E\in\mathbb{R}^{L\times L}\); (3) \textbf{Noise intensity assessment}: for each channel, we extract the mean off‐diagonal similarity at each layer index \(i\), compute distortions \(\Delta_i = 1 - x_i\), and fit \(\Delta_i\) versus \(i\) to the linear model \(\Delta_i \approx i\,\mu\,\xi_0\) (with \(\xi_0=0.5\)) via constrained least‐squares through the origin. The estimated slopes \(m\) yield channel‐specific noise intensities \(\mu = m/\xi_0\), quantifying the per‐layer accumulation rate of semantic (\(\mu_S\)) and emotional (\(\mu_E\)) perturbations.}
    \label{fig:framework}
\end{figure*}

In this section, we leverage a real‐world case study and the analytical framework introduced earlier to assess the effects of such disturbances on information dissemination. Drawing on large‐scale, multilayered forwarding logs and comment data, we perform statistical analyses of observed values to measure changes in both semantic and affective consistency between layers. This approach enables us to precisely quantify the magnitude of noise and its cumulative characteristics throughout the propagation process. Such empirical evaluation not only validates theoretical model predictions but also supplies the quantitative foundation necessary for designing more resilient synchronization mechanisms.

We emphasize that each evaluation dimension—whether pertaining to emotional transmission noise, semantic transmission noise, or normative (moral) transmission noise—is considered independently. Consequently, a comprehensive assessment framework must incorporate multidimensional metrics to construct a complete “noise‐space” index.

\begin{table*}[htbp]
\centering
\caption{Emotion prediction results for each layer.}
\label{table:emotion_pre}
\begin{tabular}{ccccccccc}
\toprule
 & Joy & Trust & Anticipation & Fear & Surprise & Sadness & Disgust & Anger \\
\midrule
Layer 0 & 0.3897 & 0.3951 & 0.0238 & 0.1031 & 0.0497 & 0.0141 & 0.0014 & 0.0238 \\
Layer 1 & 0.3897 & 0.2445 & 0.1495 & 0.0209 & 0.1042 & 0.0574 & 0.0221 & 0.0118 \\
Layer 2 & 0.2467 & 0.1495 & 0.3412 & 0.0253 & 0.0496 & 0.0254 & 0.0333 & 0.0333 \\
\bottomrule
\end{tabular}
\end{table*}

To empirically validate the theoretical framework presented in Sec.\ref{sec:result}, we conduct a hierarchical analysis using a real-world case study. Specifically, this study focuses on a Weibo post published by the official CCTV News account on December 18, 2024 \cite{weibo_data_source}. Data collection concluded on December 26, 2024, yielding a comprehensive dataset of over 1.31 million shares, 112,000 likes, and 6,940 comments—making it well suited for our empirical investigation.

Limiting the observation window to eight days (18--26 Dec 2024) was a study-specific design choice. The window was selected to cover the principal burst-and-decay phase of the observed cascade while providing a fixed endpoint for network construction and analysis. Previous studies have documented pronounced temporal dynamics in information diffusion and emphasized the importance of explicitly modeling temporal evolution in social-media diffusion \cite{matsubara2012risefall, li2017modeling, iamnitchi2023modeling}. The eight-day interval is therefore used here as an empirical observation window for this case study rather than as a universal field-standard duration.

The dataset was procured through ethically and legally compliant channels utilizing the Weibo API, encompassing the full retweet cascade tree alongside the textual content of all associated comments. To construct the DHN, we delineated network layers according to the shortest-path (hop) distance from the source node: Layer0 comprises the originating post, while a user is assigned to Layer$i$ if the minimal retweet path from the source to that user entails precisely $i$ retweets. For this analysis, we delimited our focus to the initial three layers to preserve data integrity and delineate a coherent hierarchical topology. From the raw corpus, we excised non-textual artifacts and purged spam via an integrated approach of keyword-based filtering and anomaly detection algorithms. The comprehensive data processing pipeline is depicted in Fig.~\ref{fig:framework}.

The resulting network structure is illustrated in Fig.~\ref{fig:WeiboNetwork}, with corresponding topological parameters listed in Tab.~\ref{tab:network_params}. The original post is denoted as Layer~0 (sky-blue node). Layer~1 (green nodes) comprises users who directly retweeted the source. We further categorized these users based on their network influence: nodes with a follower count exceeding 100,000 are classified as ``celebrity users'' (dark green), while others are classified as ``average users'' (light green). Layer~2 (red nodes) represents users who retweeted from any Layer~1 user, thus forming the second propagation tier. This observed structure aligns well with the hierarchical architecture formulated in our theoretical model, where high-degree celebrity nodes serve as diffusion hubs. We emphasize that the topological characteristics observed in this case, including its directed, acyclic flow and layered partitioning, satisfy all the defining features of a DHN as specified in Sec.~\ref{sec:DHN}.

\begin{figure}
    \centering
    \includegraphics[width=0.75\linewidth]{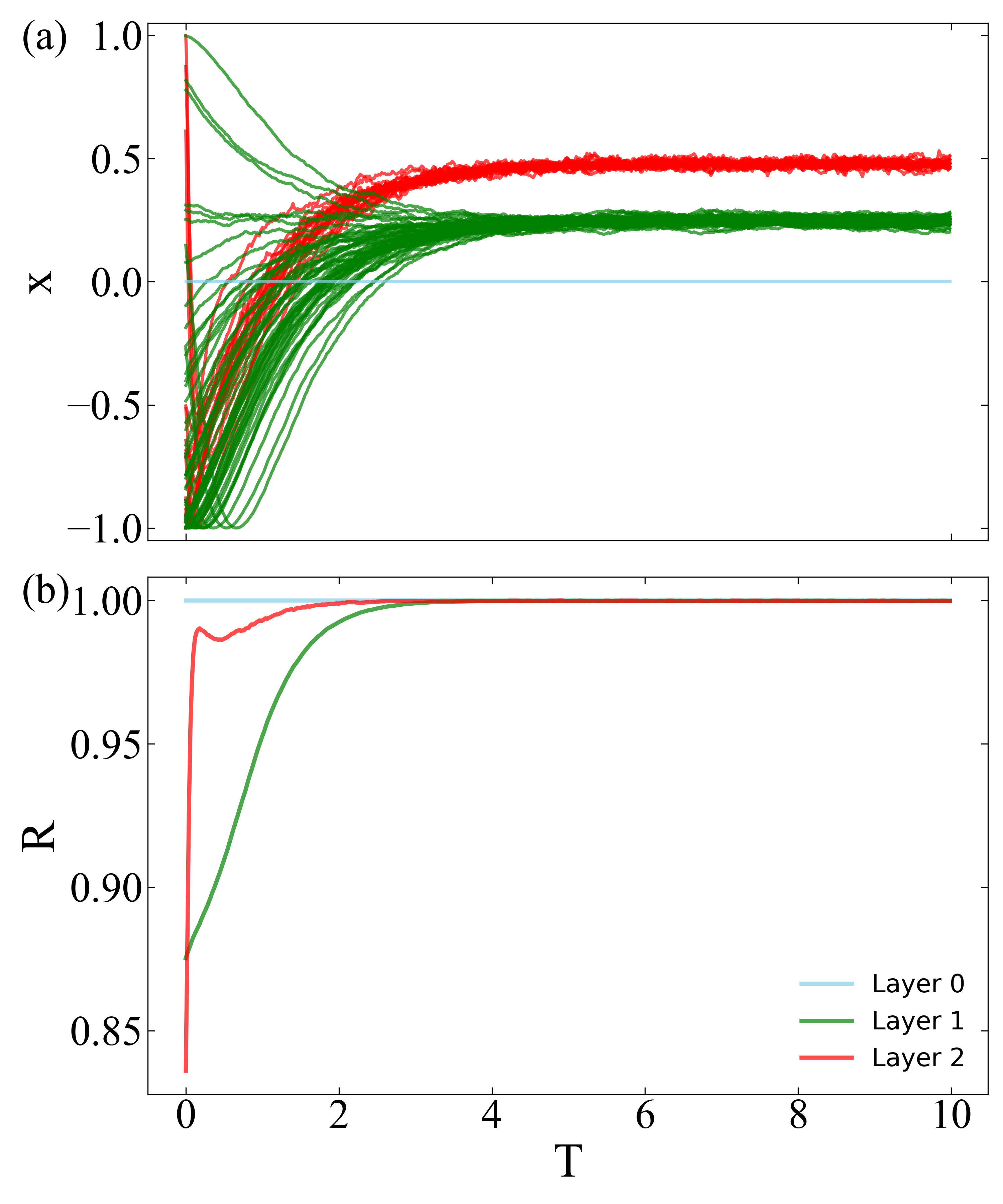}
    \caption{The dynamical behavior of NFHK in DHN shown as Fig. \ref{fig:WeiboNetwork}, with time series (a) of the system state $x$, and (b)  of the order parameter $R$. The selection of system topology structure parameters is in accordance with Tab.\ref{tab:network_params}, and the selection of dynamic parameters is $\mu = 0.5$.}
    \label{fig:Timeseries_Real}
\end{figure}

We analyzed the dynamical behavior of the DHN within this topology, as shown in Fig.~\ref{fig:Timeseries_Real}. From the system state $x$ (Fig.~\ref{fig:Timeseries_Real}(a)) and the order parameters of each layer (Fig.~\ref{fig:Timeseries_Real}(b)), we observed a clear hierarchical synchronization phenomenon in opinion dynamics. The opinion value $x_0$ of the source node remained unchanged. All opinions of the nodes in the first layer gradually converged to a viewpoint that deviated from the source node’s opinion, indicating synchronization within the first layer. The opinions of the second-layer nodes converged to a viewpoint that deviated further from the source node, signifying synchronization within the second layer. Additionally, adjacent-layer nodes exhibited nearly identical opinion values, and the range of minor perturbations in opinions within the first and second layers was nearly identical.

We present the Multi-channel NFHK framework for assessing noise intensity in hierarchical social networks in Fig.~\ref{fig:framework}. The method constructs two parallel channels—semantic and emotional—through two classical text analysis dimensions. Information is evaluated layer by layer within these channels. In the semantic channel, raw comment text is preprocessed to remove irrelevant tokens, then embedded using a Word2Vec model trained on large-scale Chinese corpora. Pairwise cosine similarities between comment embeddings yield a matrix quantifying content convergence across propagation depths. Simultaneously, the emotional channel employs a fine-tuned BERT classifier to project each comment onto an 8-dimensional Plutchik emotion wheel, with cosine similarity again measuring affective alignment. By comparing these similarity matrices across successive layers, the framework quantifies the cumulative noise intensity—i.e., the extent to which external perturbations drive unintended homogenization of meaning and sentiment along the retweet/comment chains” 

\begin{table}[ht]
    \centering
    \caption{Parameters for selected layers in the DHN shown in Fig.~\ref{fig:WeiboNetwork}.}
    \label{tab:network_params}
    \begin{tabular}{cccc}
        \toprule
        \textbf{Layer} ($i$)& $N_i$& $p_i$& $\beta_i$\\
        \midrule
        1& 583 & 1 & 58 \\
        2& 6832 & $1.17 \times 10^{-3}$ & 11.7 \\
        \bottomrule
    \end{tabular}
\end{table}

\begin{figure}[htbp]
    \centering
    \includegraphics[width=1\linewidth]{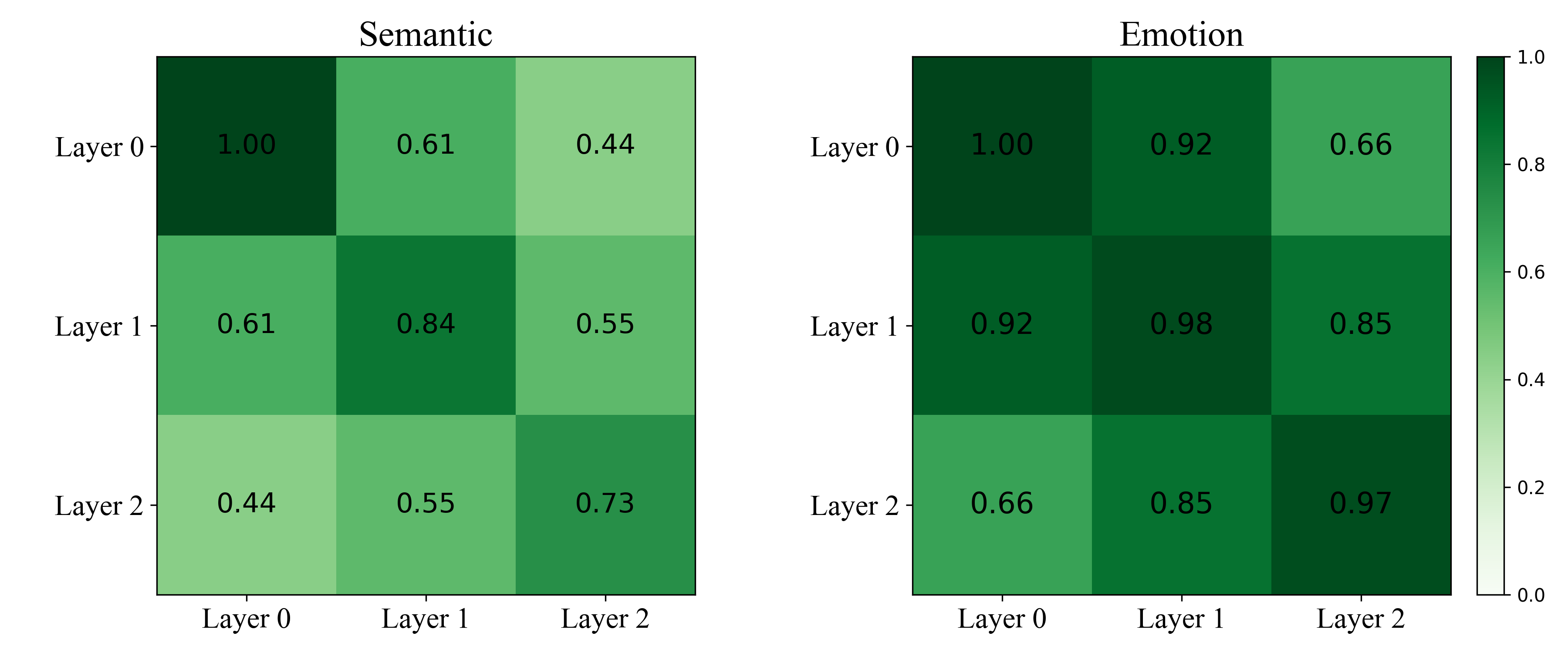}
    \caption{Similarity matrices across network layers. Left: semantic similarity. Right: emotional similarity. Darker shades indicate stronger similarity.}
    \label{fig:similar}
\end{figure}

To quantify convergence in both content and emotion across layers, we propose a dual-channel similarity analysis framework in Fig.~\ref{fig:framework}. In the semantic channel, comments were preprocessed by removing links, emojis, and user mentions, followed by Chinese word segmentation using Jieba. Word2Vec embeddings trained on a large-scale Chinese corpus were used to compute semantic similarity via cosine distance:
\begin{equation}
    CS(A,B) = \frac{A \cdot B}{\|A\| \|B\|}.
\end{equation}
In the emotional channel, a fine-tuned BERT model predicted emotion vectors based on Plutchik’s wheel of emotions, with cosine similarity applied to assess emotional alignment.

We report the average predicted probabilities for each emotion category at layers 0–2 in Tab.~\ref{table:emotion_pre}.  Notably, layer 2 displays a more uniform profile—marked by elevated anticipation and diminished trust and joy—consistent with the model’s prediction of emotional convergence during propagation.

\begin{figure}[htbp]
    \centering
    \includegraphics[width=1\linewidth]{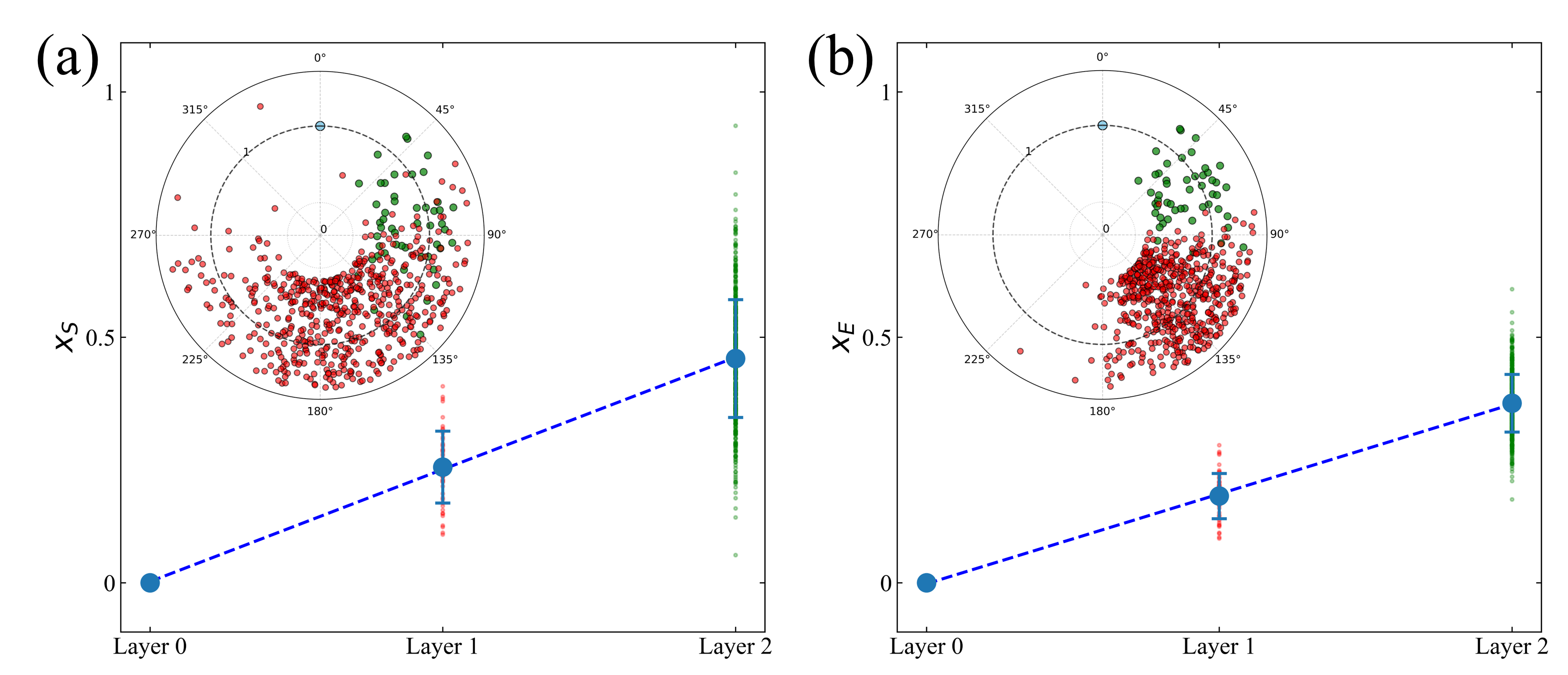}
    \caption{Layer-wise evolution of similarity. (a) Semantic similarity $x_S$; (b) Emotional similarity $x_E$. Error bars indicate standard deviation. Insets: polar plots showing clustering of information units.}
    \label{fig:similarity_evolution}
\end{figure}

We present the semantic and emotional similarity matrices in Fig.~\ref{fig:similar}, which both exhibit a monotonic increase in similarity as the network depth grows. This trend is made explicit in Fig.~\ref{fig:similarity_evolution}, where the upward trajectories and reduced layer‐wise variance quantitatively mirror the behavior predicted by Eq.~\eqref{eq:distortion_mean}.

To estimate the per‐layer noise intensity \(\mu\), we proceed as follows.  From Fig.~\ref{table:emotion_pre}, we extract the mean semantic similarities \(S_i\) and emotional similarities \(E_i\) at layers \(i = 0,1,2\), and define the distortions
\[
  \Delta_i^{(S)} = 1 - S_i,
  \quad
  \Delta_i^{(E)} = 1 - E_i.
\]
Under the fractal‐topology noise model—where distortions accumulate linearly with depth—we posit
\( \Delta_i \approx i \,\mu\,\xi_0\) with bias \(\xi_0 = 0.5\).  A constrained least‐squares fit through the origin on \(\{(i,\Delta_i)\}\) yields a slope \(m\), from which\( \mu = \frac{m}{\xi_0}.\)

Applying this procedure gives \(\mu_S \approx 0.32\) for the semantic channel and \(\mu_E \approx 0.30\) for the emotional channel, thereby quantifying the rate of noise accumulation per layer.

Together, these results furnish strong empirical support for our theoretical framework.  The concordance between observed convergence in both content and emotion and the analytic predictions underscores the self‐reinforcing dynamics of hierarchical information flow in real‐world social media networks.

\section{Discussion}
\label{sec:dis}

This section discusses the theoretical assumptions, parameter choices, and potential extensions of the proposed model, with the aim of clarifying the scope and implications of the obtained results.

\subsection{Noise Model Specification} 
In this study, a Gaussian white noise process with a nonzero mean (\(\xi_0 = 0.5\)) was introduced to represent systematic bias in social media information diffusion. This modeling choice is theoretically well-grounded and analytically convenient. From the perspective of stochastic processes and statistical physics, Gaussian noise naturally emerges as the limit of aggregated independent perturbations, consistent with the central limit theorem. Accordingly, this assumption allows the analytical derivation of key results from the stochastic differential formulation, including the linear accumulation of distortion across hierarchical layers [Eq.~\eqref{eq:layer_mean_final}] and the intra-layer synchronization variance [Eq.~\eqref{eq:width}].

The adoption of a nonzero mean is not arbitrary but reflects the directional bias frequently observed in real-world communication systems. Social media dynamics are often shaped by non-random forces such as algorithmic curation, group polarization, and dominant narrative framing, which collectively introduce a persistent shift in information flow. In this context, the mean of the noise captures the systematic bias, while its variance represents random fluctuations. Since the model output primarily depends on the product \(\mu \cdot \xi_0\), fixing \(\xi_0 = 0.5\) serves as a normalization that simplifies parameter estimation and preserves generality without loss of interpretability.

Importantly, the core findings of this model exhibit strong robustness to alternative noise distributions. Any independent and identically distributed noise with a finite nonzero mean—such as uniform or Laplacian noise—leads to the same linear distortion accumulation pattern, with the slope determined by the specific distributional characteristics. Moreover, the persistence of intra-layer synchronization mainly depends on the symmetry of the network and the linearity of its dynamics. As long as the noise remains statistically identical within each layer, the synchronized state is expected to remain stable in a statistical sense.

Future extensions may incorporate more complex stochastic mechanisms, including temporally correlated noise to capture user interdependence and algorithmic coupling, heavy-tailed noise to model rare but extreme opinion shocks, and time-varying noise intensity to reflect the evolving randomness of information diffusion. Such generalizations would provide a more realistic representation of information distortion in digital communication networks and offer finer theoretical insights for the design of platform governance and algorithmic intervention strategies.

\subsection{The Role of the Zero-Drift Assumption} 

The NFHK model assumes that all nodes possess zero intrinsic drift, \(F_j(x_j) = 0\), in order to isolate the effects of network topology and stochastic forcing. By removing internal dynamics at the node level, the model highlights how structural connectivity and external perturbations jointly determine information distortion and synchronization patterns. This simplification yields analytical tractability and establishes a theoretical baseline against which more complex behaviors can be compared.

Nevertheless, in some specific social systems, individual users typically exhibit intrinsic tendencies or external motivations, implying \(F_j(x_j) \neq 0\). Depending on whether such drift terms are homogeneous or heterogeneous across the network, the system may display qualitatively different dynamics, including phase locking, partial synchronization, or multi-stable configurations. Accounting for these factors would enrich the model’s descriptive and predictive power, offering deeper understanding of opinion formation and collective behavioral dynamics in complex communication networks.

\section{Conclusion}\label{sec:con}

This study proposes a unified framework that integrates fractal topology and stochastic dynamics to examine the structural mechanisms behind information distortion in public opinion networks. Through theoretical derivations, numerical simulations, and empirical validation, it demonstrates how seemingly chaotic opinion dynamics can be understood, predicted, and governed through quantifiable variables and targeted interventions.

At the theoretical level, the research establishes that information distortion accumulates linearly with the number of propagation layers, expressed as $\Delta_i = i\mu\xi_0$, where $\mu$ denotes the noise intensity and $\xi_0$ is the initial deviation. It identifies key topological variables such as connection probability $p$ and fractal dimension $\beta$ as primary factors shaping the evolution of distortion. The emergence of intra-layer synchronization and remote hierarchical synchronization reveals that even users without direct links can exhibit correlated behaviors, driven by the interplay between stochastic noise and hierarchical structure. These findings challenge the conventional view that opinion diffusion is primarily governed by local, direct connections, and instead highlight the importance of global topological patterns in shaping collective sentiment.

Building on these theoretical insights, the study proposes a dual-path governance strategy grounded in topology control and noise regulation. On the structural side, reducing inter-layer connection probability is shown to lower intra-layer fluctuations and suppress the accumulation of distortion. On the dynamic side, injecting reverse noise into critical layers—especially at the grassroots level—can effectively neutralize systemic bias, enabling the recovery of synchronized opinion states across layers. The model also demonstrates that increasing the coupling strength of authoritative nodes helps stabilize early-stage opinions and mitigate the spread of misinformation.

These mechanisms are not merely theoretical. The study offers practical applications in public opinion monitoring and intervention. By detecting intra-layer synchronization—such as converging sentiment patterns in user comments—and combining semantic similarity with sentiment vector analysis, the model enables early identification of polarization risks. It also introduces a distortion-based source-tracing method that leverages the linear accumulation law to efficiently backtrack the origin of rumors, as demonstrated in a case study of misinformation spread on Weibo. Layers with high connection probabilities are identified as distortion amplifiers, signaling zones that require prioritized governance.

In terms of real-world implementation, the model encourages the development of platform-level tools such as topology-aware algorithms and visualization systems. These can help automate fact-checking in high-distortion areas and improve users’ awareness of media manipulation. More broadly, the study translates abstract notions like “echo chambers” and “information cocoons” into quantifiable metrics such as $\Delta_i$ and $\sigma_i$, making them accessible for algorithmic diagnosis and policy formulation.

Nevertheless, the study acknowledges several limitations. The current model assumes a relatively static network topology, while in practice, public opinion networks evolve dynamically, especially during breaking news or emergencies. Future research should explore the extension of this framework to time-varying networks in order to capture real-time changes in distortion dynamics. Furthermore, the current model treats users as homogeneous agents, ignoring individual variations in cognitive ability, network influence, and media literacy. Introducing agent-based models with heterogeneous attributes could improve the realism and predictive power of the framework. The issue of cross-platform opinion propagation also remains underexplored; quantifying cumulative distortion effects across multiple platforms requires the construction of unified fractal topology datasets and closer collaboration between computational and social science disciplines.

Finally, this study argues for the incorporation of topological parameters and noise metrics into the regulatory frameworks used by digital platforms. By establishing industry standards around variables such as $p$, $\beta$, and $\mu$, regulators and platform designers can move toward more transparent, accountable, and scientifically grounded information governance practices. Such standards would not only help mitigate misinformation risks but also preserve open and healthy public discourse.

In conclusion, this research illustrates the power of mathematical modeling in addressing complex social challenges. By revealing the hidden order within the apparent chaos of public opinion distortion, it lays a scientific foundation for building resilient digital ecosystems—ones that balance freedom of expression with the need for structural robustness, and that use the language of networks to inform the governance of the public sphere.


\bibliography{Ref_Luo}

\appendix

\section{Derivation of Eq.\ref{eq:layer_mean_dyn_sync} }
\label{sec:derivation_eq5}
We provide here a detailed derivation of Eq.~(5) that describes the dynamics of the layer-averaged opinion under the NFHK model. Consider node $j$ in layer $i$, whose dynamics is given by
\begin{equation}
\dot{x}_{i,j} = \sum_{k \in \mathcal{N}^{\mathrm{in}}_{i,j}} \left( x_k - x_{i,j} + \mu \xi_j(t) \right),
\end{equation}
where $\xi_j(t)$ is a stochastic process satisfying $\langle \xi_j(t) \rangle = \xi_0$ and $\langle \xi_j(t) \xi_j(s) \rangle = \delta(t - s)$.

The layer-averaged opinion is defined as
\begin{equation}
\langle x_i \rangle = \frac{1}{N_i} \sum_{j=1}^{N_i} x_{i,j}.
\end{equation}
Taking the time derivative and substituting the dynamics of each node gives
\begin{equation}
\frac{d}{dt} \langle x_i \rangle = \frac{1}{N_i} \sum_{j=1}^{N_i} \sum_{k \in \mathcal{N}^{\mathrm{in}}_{i,j}} \left( x_k - x_{i,j} + \mu \xi_j(t) \right).
\end{equation}

The first term represents the sum of opinions from the in-neighbors of nodes in layer $i$, and can be rewritten as
\begin{equation}
\frac{1}{N_i} \sum_{(k,j) \in E_{i-1,i}} x_k = \frac{1}{N_i} \sum_{k \in L_{i-1}} d_k^{\mathrm{out}} x_k,
\end{equation}
where $d_k^{\mathrm{out}}$ is the out-degree of node $k$. Under the mean-field approximation that $x_k \approx \langle x_{i-1} \rangle$ for all $k$ in layer $i-1$, this term becomes
\begin{equation}
\langle x_{i-1} \rangle \frac{1}{N_i} \sum_{k} d_k^{\mathrm{out}} = \langle x_{i-1} \rangle N_{i-1} p_{i-1},
\end{equation}
where $p_{i-1}$ is the connection probability from layer $i-1$ to layer $i$.

The second term is the contribution from the nodes' own opinions, weighted by their in-degrees. This is expressed as
\begin{equation}
\frac{1}{N_i} \sum_{j} d_j^{\mathrm{in}} x_{i,j} \approx \langle d_i^{\mathrm{in}} \rangle \langle x_i \rangle,
\end{equation}
where $\langle d_i^{\mathrm{in}} \rangle = N_{i-1} p_{i-1}$ is the expected in-degree.

The final term accounts for the noise contribution:
\begin{equation}
\frac{\mu}{N_i} \sum_{j} d_j^{\mathrm{in}} \xi_j(t).
\end{equation}
Taking the expectation yields
\begin{equation}
\mathbb{E}\left[ \frac{\mu}{N_i} \sum_{j} d_j^{\mathrm{in}} \xi_j(t) \right] = \mu \langle d_i^{\mathrm{in}} \rangle \xi_0.
\end{equation}

Combining these results, we obtain
\begin{equation}
\frac{d}{dt} \langle x_i \rangle = \langle d_i^{\mathrm{in}} \rangle \left( \langle x_{i-1} \rangle - \langle x_i \rangle \right) + \mu \langle d_i^{\mathrm{in}} \rangle \xi_0,
\end{equation}
which corresponds to Eq.~(5) in the main text. This derivation highlights the reliance on the mean-field approximation, the large-network limit, and the role of nonzero noise mean $\xi_0$ in introducing systematic bias in hierarchical opinion dynamics.

\end{document}